\documentclass[twocolumn,superscriptaddress,preprintnumbers,amsmath,10pt,aps,prl,nobibnotes,longbibliography]{revtex4-1} 

\usepackage{algorithm, algorithmic}

\usepackage{amsmath}
\usepackage{amsfonts}
\usepackage{bm}
\usepackage{color}
\usepackage{verbatim}
\usepackage{graphicx}
\usepackage{lipsum}
\usepackage[utf8]{inputenc}
\usepackage{times}
\usepackage{graphicx}
\usepackage{bm}
\usepackage{amssymb,multirow}
\usepackage[dvipsnames]{xcolor}
\usepackage[normalem]{ulem} 
\usepackage[mathlines]{lineno}
\usepackage{bbm}
\usepackage{siunitx}
\usepackage{float}
\usepackage{siunitx}

\def\Vect{\mathbf{t}}
\def\Vecu{\mathbf{u}}

\def\VecA{\mathbf{A}}

\def\VecC{\mathbf{C}}
\def\VecD{\mathbf{D}}

\def\VecH{\mathbf{H}}
\def\VecI{\mathbf{I}}

\def\VecK{\mathbf{K}}
\def\VecL{\mathbf{L}}
\def\VecM{\mathbf{M}}

\def\VecP{\mathbf{P}}

\def\VecS{\mathbf{S}}
\def\VecT{\mathbf{T}}

\usepackage{soul}
\usepackage[size=scriptsize]{todonotes}

\begin{document}

\title{How to build the `optical inverse' of a multimode fibre}

\author{Un\.e~G.~B\=utait\.e}
\thanks{These authors contributed equally to this work.}
\affiliation{School of Physics and Astronomy, University of Exeter, Exeter, EX4 4QL. UK.}
\email{u.butaite@exeter.ac.uk}
\author{Hlib Kupianskyi}
\thanks{These authors contributed equally to this work.}
\affiliation{School of Physics and Astronomy, University of Exeter, Exeter, EX4 4QL. UK.}
\author{Tomáš Čižmár}
\affiliation{Leibniz Institute of Photonic Technology, Albert-Einstein-Straße 9, 07745 Jena, Germany.}
\affiliation{Institute of Scientific Instruments of the CAS, Královopolská 147, 612 64 Brno, Czech Republic.}
\author{David~B.~Phillips}
\email{d.phillips@exeter.ac.uk}
\affiliation{School of Physics and Astronomy, University of Exeter, Exeter, EX4 4QL. UK.}

\begin{abstract}
When light propagates through a multimode optical fibre (MMF), the spatial information it carries is scrambled. Wavefront shaping can undo this scrambling, typically one spatial mode at a time -- enabling deployment of MMFs as ultra-thin micro-endoscopes. In this work we go beyond serial wavefront shaping by showing how to simultaneously unscramble all spatial modes emerging from an MMF in parallel. We introduce a passive multiple-scattering element -- crafted through the process of inverse design -- that is complementary to an MMF and undoes its optical effects. This `optical inverter’ makes possible both single-shot wide-field imaging and super-resolution imaging through MMFs. Our design consists of a cascade of diffractive elements, and can be understood from the perspective of both {\it multi-plane light conversion}, and as a physically inspired {\it deep diffractive neural network}. This physical architecture can outperform state-of-the-art electronic neural networks tasked with unscrambling light, as it preserves the phase and coherence information of the optical signals flowing through it. Here we demonstrate our MMF inversion concept through numerical simulations, and efficiently sort and unscramble up to $\sim400$ step-index fibre modes, reforming incoherent images of scenes at arbitrary distances from the distal fibre facet. We also describe how our optical inverter can dynamically adapt to see through flexible fibres with a range of experimentally realistic TMs -- made possible by moulding optical memory effects into the structure of our design. Although complex, our inversion scheme is based on current fabrication technology so could be realised in the near future. Beyond imaging through scattering media, these concepts open up a range of new avenues for optical multiplexing, communications and computation in the realms of classical and quantum photonics.
\end{abstract}

\maketitle

\noindent{\bf Introduction}\\
Multimode optical fibres (MMFs) are a ubiquitous technology in light guiding applications. They form the backbone of the short-range optical links in data centres, transmit optical signals in spectrometers, and deliver high intensity light in laser welding systems~\cite{singh2015jupiter}. MMFs support thousands of spatial modes -- i.e.\ unique optical field profiles -- within the width of a strand of hair, each mode representing an independent communication channel. There has been much interest in unlocking this high spatial information density for imaging and optical communications applications~\cite{kahn2017communications}. For instance, recently MMFs have emerged as a promising technology for micro-endoscopy -- enabling imaging with sub-cellular resolution at the tip of a needle~\cite{turtaev2018high,ohayon2018minimally}. In this case, each spatial mode is capable of conveying a different subset of the spatial information from a scene at the distal (far) end of the fibre. Unfortunately, spatial mode dispersion scrambles this information as it propagates along an MMF. Therefore optical signals must be unscrambled at the proximal (near) end in order to reconstruct images~\cite{spitz1967transmission,choi2012scanner,vcivzmar2012exploiting,papadopoulos2012focusing}.

The current `gold-standard' method to overcome optical scrambling is via measurement of the fibre's transmission matrix (TM): a linear matrix operator that describes how any monochromatic field will have been transformed upon propagating through the fibre~\cite{popoff2010measuring}. Once measured, the TM can be used to calculate a sequence of pre-scrambled input fields which are sent from the proximal end to generate a scanning focus at the distal facet. Thus scanning imaging is possible by correlating the return signal with each known focussed scan location. This method, known as wavefront shaping~\cite{vellekoop2007focusing,mosk2012controlling,gigan2021roadmap}, has been used to look through MMFs centimetres deep inside living tissue~\cite{turtaev2018high,ohayon2018minimally} -- a task difficult to accomplish in any other such minimally invasive way. A wide variety of imaging modalities have been demonstrated through these MMF-based holographic endoscopes, including fluorescence imaging~\cite{vcivzmar2012exploiting,turtaev2018high,ohayon2018minimally}, chemically and structurally sensitive microscopy~\cite{gusachenko2017raman,tragaardh2019label,cifuentes2021polarization}, imaging beyond the distal fibre facet~\cite{leite2021observing}, and depth ranging using time-of-flight LiDAR~\cite{stellinga2021time}.

\begin{figure*}[t]
   \includegraphics[width=1\textwidth]{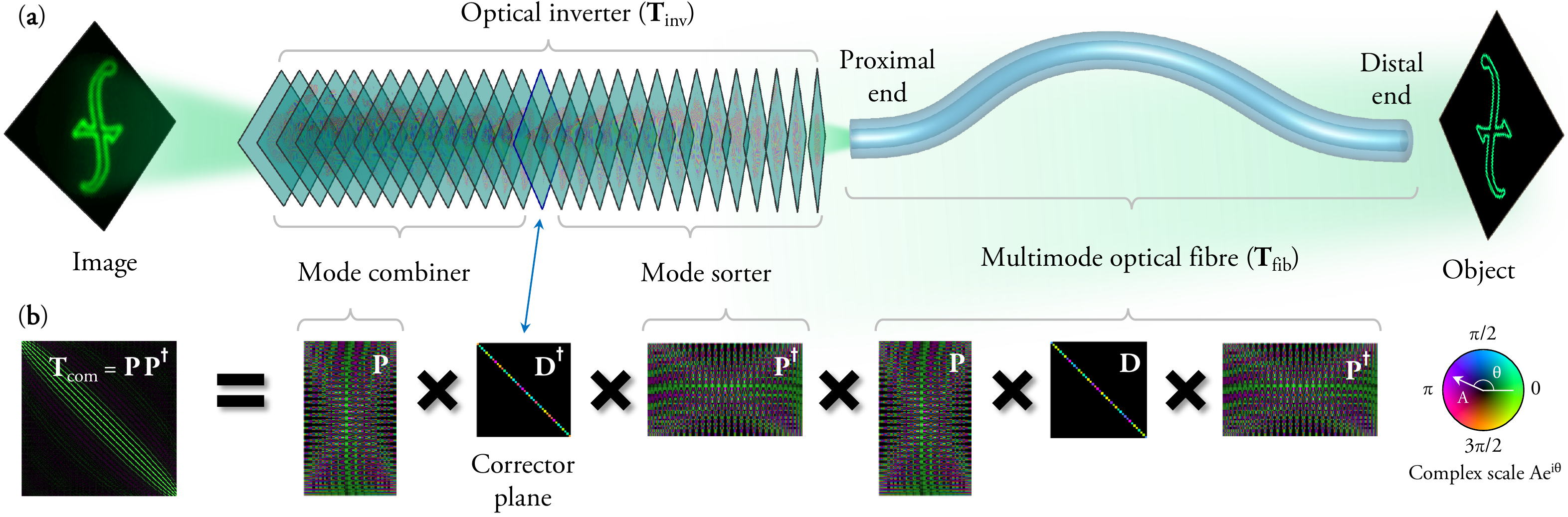}
   \caption{{\bf Optical inversion concept}. ({\bf a}) A schematic showing all-optical image reformation of a scene at the distal end of an MMF. Light emanating from the scene propagates through an MMF, thus scrambling its spatial information. The light then propagates through an optical inverter consisting of a cascade of 29 optimised phase masks, each separated by free-space. At the output of the optical inverter an image of the scene is reformed. Its resolution is governed by the spatial frequencies supported by the MMF. ({\bf b}) The action of an ideal MMF and optical inverter represented as complex matrix operators. The inverter is designed to undo the fibre mode dependent phase delays imparted by propagation through the MMF. The combined transmission matrix of the MMF and inverter, $\VecT_{\mathrm{com}}$, is equivalent to the identity matrix convolved with the diffraction limited point-spread function dictated by the numerical aperture of the MMF.}
   \label{Fig:schematic}
\end{figure*}

The majority of TM-based imaging techniques demonstrated so far rely on raster-scanning or sequential pattern projection, essentially meaning that light is unscrambled one spatial mode at a time~\cite{gigan2021roadmap}. This currently precludes the delivery of wide-field imaging techniques through MMFs. For example, there is no way to conduct super-resolution imaging modalities such as photo-activated localisation microscopy (PALM)~\cite{betzig2006imaging}, or stochastic optical reconstruction microscopy (STORM)~\cite{rust2006sub} at the tip of a MMF. Scanning also places limits on the maximum imaging frame-rate, which is governed by the update rate of the spatial light modulator (SLM) used to shape input optical fields.

Here we propose a new imaging strategy that goes beyond serial wavefront shaping, and simultaneously unscrambles all spatial modes emerging from a fibre in parallel. To achieve this we describe the design of a passive optical element
-– which we term an {\it optical inverter} –- that, in the ideal case, possesses a TM exactly equivalent to the inverse of the MMF’s TM ($\VecD$). Coupled with the MMF, the TM of the entire system collapses to the identity matrix $\VecI$, since $\VecD^{-1}\VecD = \VecI$. Therefore, any coherent, partially-coherent or incoherent field at the distal end of the fibre, within the spatial bandwidth of the MMF and the spectral bandwidth of the combined fibre-inverter system, will be reconstructed in an all-optical manner at the output of the optical inverter. This advance brings closer the vision of channeling any form of wide-field microscopy to the tip of an MMF endoscope positioned deep inside living tissue. Figure~\ref{Fig:schematic} shows a schematic of our concept.

The prospect of building a device to passively unscramble light that has propagated through an MMF was first considered by Gover et al.\ as long ago as 1976~\cite{gover1976direct}. Yet only recently has our understanding of the inverse design of multimodal photonic systems matured to a level that renders this concept feasible. In 2012, Čižmár et al.\ took a step forward by demonstrating that a single multiplexed hologram can be used to unscramble up to $\sim$100 spatial modes that have propagated through an MMF~\cite{vcivzmar2012exploiting}. However, this approach requires post-processing to reconstruct images, and suffers from exceedingly low conversion efficiency which has restricted its broader application~\cite{vcivzmar2012exploiting,mazilu2017modal,wang2017programmable}.

In the following, we first describe the working principle of an optical inverter in the case of an ideal MMF, followed by the specifics of our design.
We numerically simulate the performance of a system capable of efficiently unscrambling up to $\sim400$ spatial modes. We next show how this concept can be extended to the more experimentally realistic fibres, leveraging recently discovered fibre memory effects~\cite{amitonova2015rotational,li2021memory}. Finally, we describe how our inverter may be designed to dynamically adapt to the changing TM of a moving fibre.

\noindent{\bf The architecture of an optical inverter}\\
MMFs can be considered a form of opaque scattering media dominated by forward scattering: the vast majority of incident light is transmitted and levels of back-reflection are very low. Therefore here we focus solely on inversion of the TM of a MMF rather than considering its full scattering matrix. We begin by analysing the structure of the TM of an ideal MMF. Solving the monochromatic wave equation in cylindrical coordinates yields a set of $N$ circularly polarised {\it propagation invariant modes} (PIMs, also referred to here as `fibre modes'). These are orthonormal eigenmodes of the fibre, maintaining the same field profile as they propagate along it without coupling to one another. Each PIM has a mode-dependent phase velocity $\beta_n$ (where $n$ indexes the mode), and so mode $n$ picks up a global phase delay of $\theta_n = \beta_n L$ upon reaching the output of a fibre of length $L$. The de-phasing of the PIMs causes their interference pattern at the output to be different from that at the input, resulting in the spatial scrambling of optical signals. 
Therefore, the TM of an MMF, ${{\VecT_{\mathrm{fib}} \in \mathbb{C}^{P \times P}}}$, represented in the real-space basis, is given by ${\VecT_{\mathrm{fib}} = \VecP\VecD\VecP^\dagger}$ -- see Fig.~\ref{Fig:schematic}(b). Here ${\VecD \in \mathbb{C}^{N \times N}}$ is a unitary diagonal matrix capturing the phase delay experienced by all $N$ PIMs supported by the fibre, ${\VecP \in \mathbb{C}^{P \times N}}$ is a matrix transforming from the $N$-dimensional PIM-space to the $P$-dimensional pixelated real-space basis, and $^{\dagger}$ represents the conjugate transpose operator. Refs.~\cite{ploschner2015seeing,li2021memory} provide details of how matrix $\VecP$ may be constructed.

To unscramble monochromatic light that has propagated through an ideal MMF, we must undo the modal phase delays picked up during transit, by imparting a mode-dependent phase correction of ${\phi_n = -\theta_n = -\beta_n L}$ to each PIM. Thus in matrix form, the TM of the optical inverter, also represented in real-space, should correspond to
\begin{equation}\label{Eqn:inverter}
    \VecT_{\mathrm{inv}} = \VecP\VecD^\dagger\VecP^\dagger,
\end{equation}
such that the TM of the combined fibre-inverter system ${\VecT_{\mathrm{com}} \in \mathbb{C}^{P \times P}}$ is given by ${\VecT_{\mathrm{com}} = \VecT_{\mathrm{inv}}\VecT_{\mathrm{fib}}=\VecP\VecP^\dagger}$, as shown in Fig.~\ref{Fig:schematic}(b). $\VecT_{\mathrm{com}}$ is approximately diagonal, but does not equal the identity matrix because it also captures the spatial filtering due to the fibre itself -- i.e.\ only the components of the field emanating from the object that are supported by the fibre can propagate through the system. However when transformed to PIM-space, the combined system does indeed collapse to the identity matrix: ${\VecP^\dagger\VecT_{\mathrm{com}}\VecP = \VecI}$. 

Optical fields emerging from a fibre are readily addressable in real-space -- it is straightforward to impart a spatially dependent phase delay directly to these fields. However, an optical inverter must act {\it non-locally} on each PIM independently -- and so physically accessing PIM-space is the key challenge to overcome in our inverter design. We achieve this by showing how to construct optical systems that perform each of the three matrix transformations on the right-hand-side of Eqn.~\ref{Eqn:inverter}.

First, light emerging from an MMF is decomposed into fibre modes -- this can be accomplished using a passive `fibre mode sorter', the novel design of which we describe below, that redirects and focuses the energy carried by each PIM into a different spatial location across a transverse plane. This step is equivalent to the transformation enacted by matrix $\VecP^\dagger$ in Eqn.~\ref{Eqn:inverter}. Next, the light from each sorted PIM passes through a single optical element, which we call the `corrector plane', that imparts mode-dependent phase delays to undo those accumulated during transit through the fibre. This step achieves the diagonal matrix operation $\VecD^\dagger$ in Eqn.~\ref{Eqn:inverter}. Finally, the fibre modes are spatially recombined by propagating through a second fibre mode sorter in the reverse direction, enacting matrix transformation $\VecP$ in Eqn.~\ref{Eqn:inverter}. Henceforth, we refer to this reverse mode sorter as a `fibre mode combiner'.
\begin{figure*}[t]
   \includegraphics[width=0.95\textwidth]{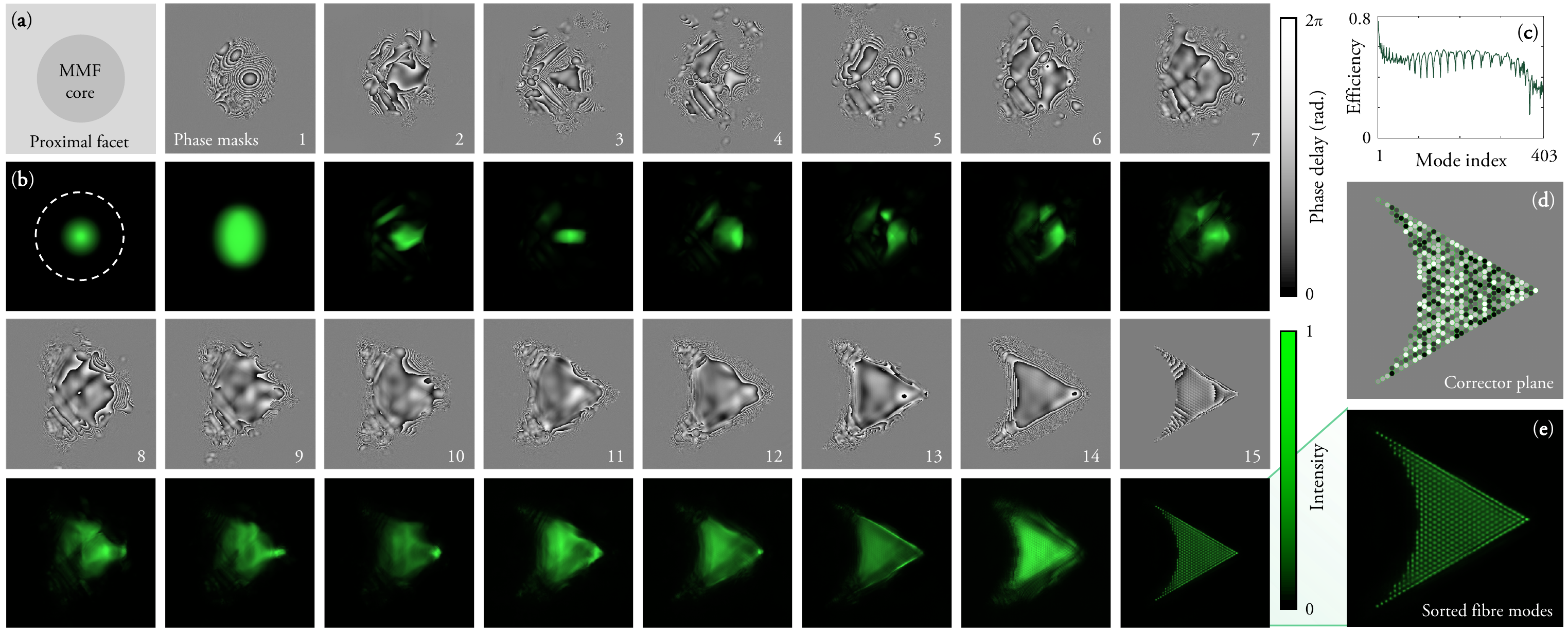}
   \caption{{\bf A 403 step-index fibre mode sorter}. ({\bf a}) The optimised phase delay profiles of the 15-phase mask MPLC. We found profiles without vertical mirror symmetry gave slightly improved performance compared to solutions where mirror symmetry was enforced. Here we use a non-cyclic colourmap to highlight the location of $0-2\pi$ phase wraps. ({\bf b}) The sum of the intensity of all 403 fibre modes as they propagate through the MPLC to separate output channels: a triangular grid of focussed Gaussian beams. Throughout the device, the mean intensity of the ensemble of fibre modes is concentrated to `smooth' regions of the phase masks with low spatial phase gradients and few phase wrapping lines. These aspects of the design will ease the experimental implementation of this MPLC. ({\bf c}) The sorting efficiency as a function of fibre mode index -- i.e.\ fraction of power of input PIM that is found in the target output channel, simulated assuming the device is unitary. See ref.~\cite{li2021memory} for details of how the PIMs are ordered. High order modes are generally sorted with a slightly lower efficiency, which could be rectified in future designs if necessary. ({\bf d}) The phase corrector plane. The location of the 403 output channels are indicated with green circles. The energy carried by each sorted PIM passes through a different region of the plane, and the phase delay imparted to each output channel can be independently modulated. ({\bf e}) A magnified view of the sorted PIMs arriving at the corrector plane -- the grid of output Gaussian spots is clearly visible.}
   \label{Fig:mode-sorter}
\end{figure*}

The number of PIMs of one circular polarisation supported by an MMF is related to its geometry, and given by ${N\sim(\pi a\mathrm{NA}/\lambda)^2}$, where $a$ is the radius of the core, $\mathrm{NA}$ is the numerical aperture of the fibre, and $\lambda$ is the wavelength of transmitted light. $N$ also governs the number of independent pixels in images that can be transmitted through an MMF. Therefore, we must design a fibre mode sorter capable of handling as many modes as there are pixels in images we wish to reconstruct.\\

\noindent{\bf A step-index fibre mode sorter}\\
High-dimensional spatial mode sorting is a challenging task, with techniques to achieve it still in their infancy. It is closely related to the inverse design of multimodal photonic systems -- an emerging class of devices capable of enacting arbitrary transformations on multiple spatial light modes simultaneously. A variety of methods to achieve this are currently under development, including meshes of waveguides on chip~\cite{reck1994experimental,carolan2015universal}, or the construction of bespoke scattering structures~\cite{bogaerts2020programmable,horodynski2022customized}, but few approaches are able to manipulate the numbers of modes required for imaging applications~\cite{molesky2018inverse}. Here we take inspiration from recent advances in {\it multi-plane light conversion}~\cite{morizur2010programmable} and {\it deep diffractive neural networks}~\cite{lin2018all}. Both of these concepts rely on a cascade of numerically optimised diffractive optical elements (DOEs, also referred to here as `phase masks') separated by free-space, to perform targeted optical transformations such as mode sorting~\cite{wang2018dynamic,fontaine2019laguerre,brandt2020high}.

Recently, Fontaine et al.\ demonstrated a multi-plane light converter (MPLC) capable of sorting hundreds of Laguerre-Gaussian (LG) modes using very few ($\sim7$) phase masks~\cite{fontaine2019laguerre,fontaine2021hermite}. The design exploits underlying symmetries of the Hermite-Gaussian (HG) basis to efficiently sort HG modes. This HG mode sorter is coupled with a cylindrical mode converter to transform from the HG basis to the LG basis~\cite{courtial1999performance}. Both LG-modes and fibre PIMs possess rotational symmetry, and this structural similarity hints that it may also be possible to sort a large number of step-index PIMs using only a small number of planes. Here we explore this possibility to design a high-dimensional step-index fibre mode sorter.

We first perform a basis transformation using a cylindrical mode converter consisting of two cylindrical lenses. We next optimise the phase profiles of a cascade of $M = 15$ phase masks, tasking the MPLC to simultaneously sort up to 403 step-index fibre PIMs. The phase profiles of these DOEs are designed using the wavefront matching method~\cite{hashimoto2005optical}, which is a form of adjoint optimisation~\cite{miller2012photonic}, and closely related to the `backpropagation' algorithms used to train electronic neural networks~\cite{werbos1990backpropagation,barre2021inverse}. This algorithm allows fast parallelised calculation of how the phase delay imparted by each pixel on a plane should be adjusted to improve the sorting performance. For more details see the SI \S1.

Figure~\ref{Fig:mode-sorter} shows our step-index fibre mode sorter design. Once the profiles of the phase masks are optimised, we do indeed find that the design inherits some of the properties of the MPLC-based LG mode sorter it is inspired by, and operates with relatively high efficiency and low modal cross-talk. We highlight that our design requires far fewer than the $\sim2N$ phase masks typically required to efficiently sort $N$ modes using multi-plane light conversion.

\begin{figure*}[t]
   \includegraphics[width=14cm]{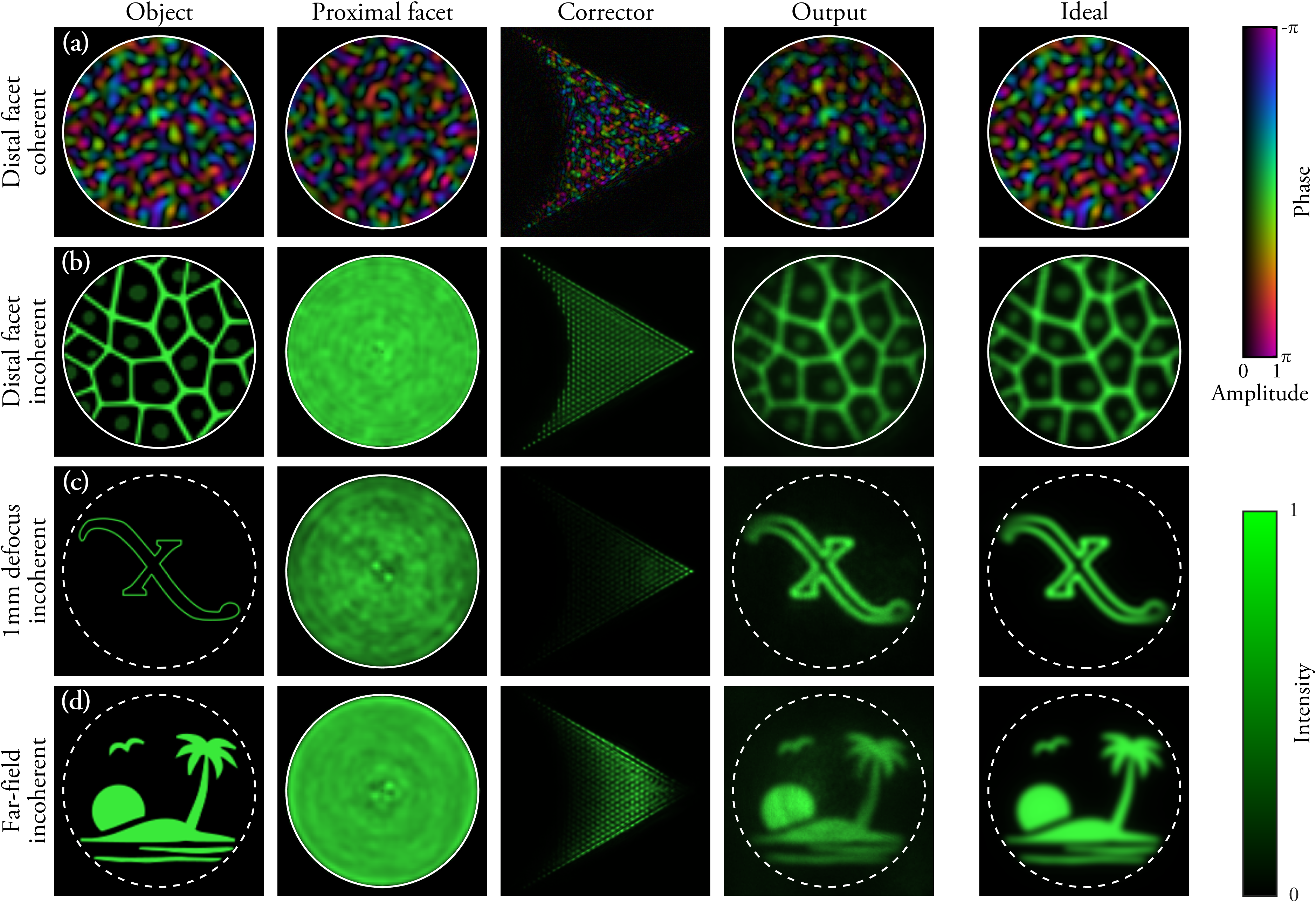}
   \caption{{\bf Seeing through ideal MMFs}. From left to right, the panels in each row show the field/object/scene to be imaged (left most panel); the field at the proximal fibre facet; the field at the corrector plane; and the reformed image at the output of our MPLC-based inverter. The right-hand panel shows the image formed using perfect MMF inversion, i.e.\ including the spatial filtering effects due to the limited modal capacity of the fibre, but not including any mode-coupling effects in the inverter. Imaging, ({\bf a}) a spatially coherent speckle pattern at the distal fibre facet; ({\bf b}) an incoherent object at the distal fibre facet; ({\bf c}) an incoherent object defocussed 1\,mm from the distal fibre facet; and ({\bf d}) in the far-field of the distal facet.}
   \label{Fig:defocus}
\end{figure*}

A unique set of design constraints can be applied to mode sorters used for imaging applications: in our case we aim to maximise the mode capacity $N$ since it is directly proportional to the pixel resolution of transmitted images. This can be achieved at the expense of a reduction in sorting efficiency, and we can also tolerate some degree of modal cross-talk, which will reduce the contrast and fidelity of transmitted images. Exploiting these trade-offs, we design a device capable of sorting $N=403$ fibre modes using 15-planes with a mean sorting efficiency of $\sim$51\%. Fig.~\ref{Fig:mode-sorter}(c) shows the mode dependent loss. We allow a mean cross-talk of $\sim$23\% -- i.e.\ on average, $\sim$23\% of the power in a given input PIM will be spread amongst the other output channels -- see SI \S1 for the characteristic coupling matrix. As shown later (see Fig.~\ref{Fig:defocus}), these levels of coupling and mode dependent loss only result in a mild reduction in imaging contrast when the full optical inverter is implemented. In addition to this 15-plane 403 fibre mode sorter, we also designed an 11-plane 201 mode sorter, and a 7-plane 107 mode sorter, which exhibit slightly improved performance due to their marginally more favourable ratios of $N/M$. SI \S1 describes the design and performance of these fibre mode sorters in detail. In each case, the output channels are arranged in a triangular grid (see Fig.~\ref{Fig:mode-sorter}(e)), chosen following ref.~\cite{fontaine2019laguerre} to help the optimisation procedure converge on a high-performance solution.\\

\noindent{\bf Looking through ideal MMFs}\\
Equipped with a design for a high-dimensional fibre mode sorter, next we numerically simulate its integration into an optical inverter to unscramble light that has propagated through an MMF. Figure~\ref{Fig:mode-sorter}(d) shows an example of a pattern of mode-dependent phase shifts applied at the corrector plane. Our 403-mode inverter requires ${2M-1 = 29}$ phase masks in total, since the last plane of the mode sorter, the corrector plane, and the first plane of the mode combiner are all in image planes of one another, so can be coalesced into a single plane. The real-space coherent TM of our design, ${\VecT'_{\mathrm{inv}} \in \mathbb{C}^{P \times P}}$ (the apostrophe distinguishing from the TM of the ideal inverter $\VecT_{\mathrm{inv}}$), is given by
\begin{equation}\label{Eqn:MPLCRealTM}
    \VecT'_{\mathrm{inv}} = \underbrace{\left[\VecL^\top\VecM^\top\right]}_\text{combiner}\hspace{1mm}\times\underbrace{\VecC}_\text{corrector}\times\hspace{1mm}\underbrace{\left[\VecM\VecL\right]}_\text{sorter}.
\end{equation}
Here ${\VecL \in \mathbb{C}^{P \times P}}$ is the TM of the optical system guiding light from the proximal fibre facet, through the cylindrical mode converter, to the input plane of the MPLC. ${\VecM \in \mathbb{C}^{P \times P}}$ is the TM of the MPLC. ${\VecC \in \mathbb{C}^{P \times P}}$ is a diagonal unitary matrix representing the TM of the corrector plane, which applies the phase correction $\phi_n$ to the area that light from the $n^{\text{th}}$ sorted PIM passes through, along with additional phase corrections to account for mode dependent phase delays imparted to the light during transit through the sorter and combiner themselves (see SI \S1 for details). $\VecM^{\top}$ represents the propagation through the MPLC in the reverse direction, where $^{\top}$ is the transpose operator. $\VecM$ is given by
\begin{equation}\label{Eqn:MPLC}
    \VecM = \VecA_M\overset{\curvearrowleft}{\prod_{m=1}^{M-1}}\left(\VecH\VecA_m\right),
\end{equation}
where $\VecA_m \in \mathbb{C}^{P \times P}$ is a unitary diagonal matrix, whose diagonal elements hold the phase change imparted by each pixel of the $m^{\mathrm{th}}$ plane. The overset arrow indicates left matrix multiplication, and $\VecH \in \mathbb{C}^{P \times P}$ is a free-space propagation matrix transforming the field from the output facet immediately after plane $m$ to the input facet immediately before plane $m+1$. See SI \S2 for more details and a description of how matrices $\VecL$, $\VecC$ and $\VecH$ are constructed. Our numerical simulation captures any reduction in inverter performance caused by the residual levels of mode-dependent loss and cross-talk exhibited by the optimised mode sorter and combiner. To simulate incoherent imaging (within the spectral bandwidth of the system), we calculate the real positive-valued incoherent intensity TM ${\VecS \in \mathbb{R}_+^{P \times P}}$, linking spatially incoherent fields at either end of the combined fibre-inverter system in the real-space basis~\cite{boniface2020non,ruan2021optical}: ${\VecS =|\VecT'_{\mathrm{inv}}\VecT_{\mathrm{fib}}|^{2}}$, where we take the element-wise square.

Thus far, MPLCs have been built from a multi-pass cavity consisting of a mirror placed opposite a reflective liquid-crystal SLM or a lithographically etched reflective DOE~\cite{fontaine2019laguerre,mounaix2020time}. Using reflective DOEs in this way leads to a mainly forward-scattering device, which matches the forward scattering nature of the MMF itself, and avoids noise associated with multiple reflections between planes that would arise in cascades of transmissive DOEs. Here we envisage that the fibre mode sorter and combiner are constructed from fixed reflective DOEs to promote high efficiency operation. For example, Fontaine et al.\ recently implemented a 14-plane MPLC composed of a gold-coated diffraction grating opposite a dielectric mirror, that exhibited a loss of ${\sim5-6\,\mathrm{dB}}$~\cite{fontaine2017design,fontaine2018packaged}. We anticipate implementing the corrector plane with a liquid-crystal SLM, which allows the modal correction phases to be dynamically reconfigured. This way, the same optical inverter can be paired with MMFs of standardised $\mathrm{NA}$ and core diameter, but of variable length, by simply modifying the correction phases. The ability to reconfigure these correction phases also provides a number of additional benefits when imaging through imperfect or flexible fibres, as will be discussed later. SI \S3 shows schematics of the full optical system of our proposed MMF inverter.

We model a step-index MMF of core radius $a=$ \SI{40.3}{\micro\meter}, and $\mathrm{NA}=0.1$, chosen to support $N = 403$ modes per polarisation at a wavelength of $\lambda=633$\,nm. We note that our optical inverter design protocol is not sensitive to the exact fibre geometry -- the ratio $N/M$ is the most critical performance predictor.

Supplementary Movie 1 shows the simulated light field at several stages through the MMF-inverter system as a focussed spot is scanned over the distal facet of the fibre. We see that high-contrast spots are reconstructed at the output of the inverter, reproducing the input spots with high fidelity. This can be quantified by measuring the spot power-ratio, which is defined as the ratio of the power within a small disk centred on the target spot position at the output plane of the inverter, to the total power transmitted within the image of the fibre core~\cite{li2021memory}. SI \S4 shows a power-ratio map, indicating the fidelity with which spots are imaged at different regions across the fibre core (see also insets in the left-most panels of Figs.~\ref{Fig:imperfect}(a-c)). Averaging the power-ratio over all output locations yields the mean power-ratio $p_{\text{r}}$. When our inverter is matched with an ideal fibre, the mean power-ratio of re-imaged points $p_{\text{r}} = 0.74$, a value comparable with the typical power-ratio for state-of-the-art serial wavefront shaping through MMFs~\cite{leite2021observing,stellinga2021time}. 

Figure~\ref{Fig:defocus} shows example images of distal scenes reconstructed at the output of the inverter system. Figures~\ref{Fig:defocus}(a-b) show images of scenes at the distal facet: a spatially coherent speckle pattern (a) and a spatially incoherent scene (b). An advantage of our concept is that refocussing of the image plane away from the end of the distal fibre facet can be achieved by simply axially re-positioning a camera recording the image at the output of the inverter. Figures~\ref{Fig:defocus}(c-d) shows the incoherent imaging of scenes located 1\,mm beyond the distal fibre facet (c), and in the far-field of the distal facet (d)~\cite{leite2021observing}. We see that in all cases the image reconstruction quality is close to the theoretical optimum for wide-field imaging (shown in Fig.~\ref{Fig:defocus}, right-hand column) in terms of both resolution and contrast. Far-field imaging exhibits the lowest fidelity, due to the greater proportion of high order PIMs excited in this configuration, which are marginally less efficiently sorted.\\
\begin{figure}[b]
   \includegraphics[width=7.5cm]{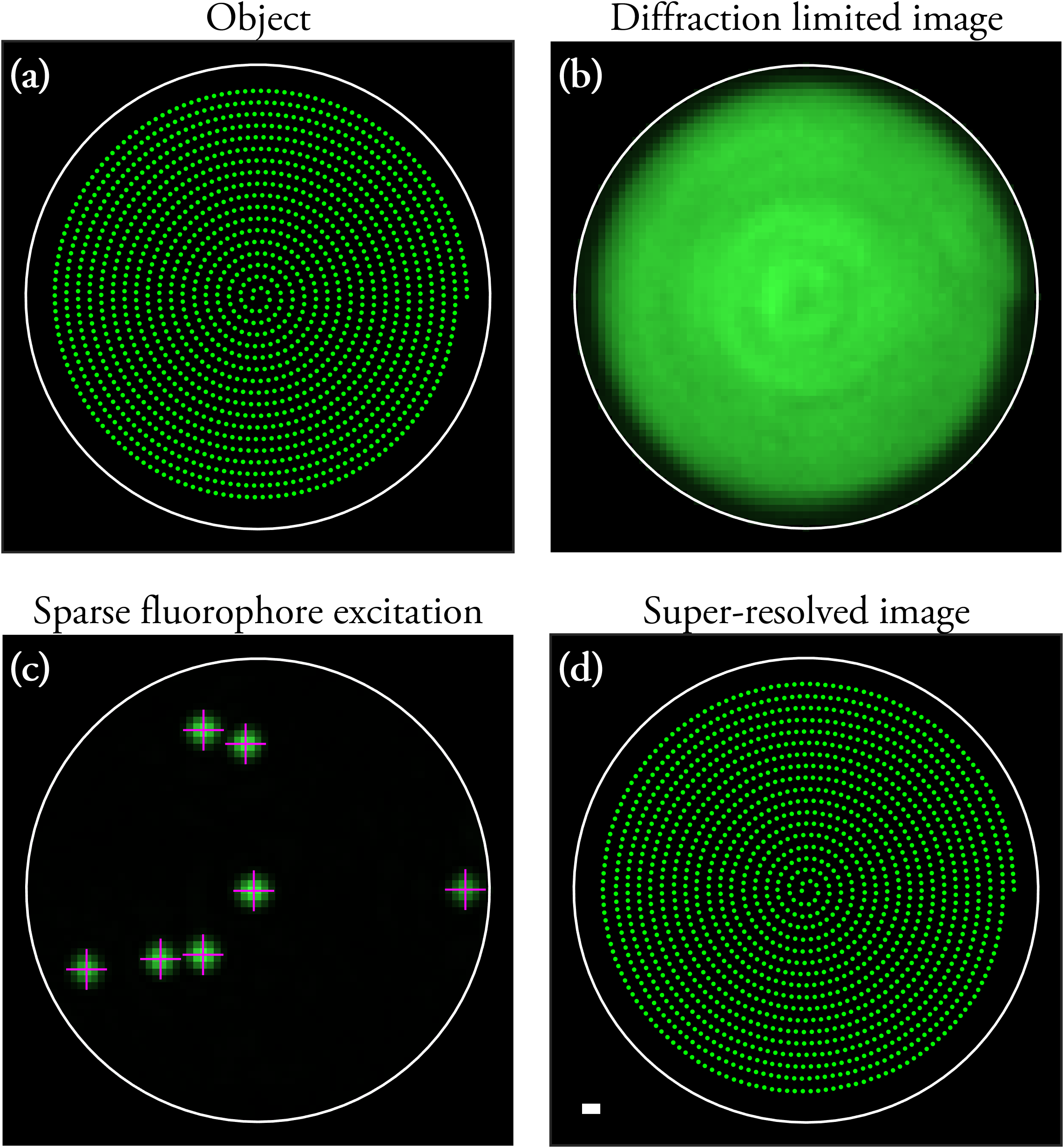}
   \caption{{\bf Super-resolution imaging}. ({\bf a}) A spiral-shaped object labelled with fluorophores. ({\bf b}) A single-shot wide-field image that cannot resolve individual fluorophores. ({\bf c}) An image of several activated fluorophores. Pink crosses show the tracked centroid positions. ({\bf d}) A super-resolved image compiled from multiple images of randomly sparsely excited fluorophores. The scale bar shows the length of the diffraction limit, which in this case is \SI{3.16}{\micro\meter}. The colour-bar for intensity images {\bf (b,c)} is the same as that in Fig.~\ref{Fig:defocus}.}
   \label{Fig:supRes}
\end{figure}

\noindent{\bf Super-resolution imaging}\\
Our inverter gives direct access to the field at the distal end of a fibre, potentially making possible a host of advanced microscopy techniques beyond wide-field imaging. For example, we now test whether the fidelity of the inversion is compatible with super-resolution imaging through a MMF -- see Fig.\ \ref{Fig:supRes}. Here we simulate PALM or STORM -- which differ only by the fluorophore switching mechanism~\cite{rust2006sub,betzig2006imaging}. These techniques are based on the principle that it is possible to locate the centroid of a diffraction limited spot to a better precision than the diffraction limit itself. Multiple wide-field fluorescent images are recorded, with fluorophore excitation carefully controlled so that only a random well-separated subset of fluorophores are induced to emit light in any given image. Centroid tracking of the isolated diffraction limited spots then reveals the position of each fluorophore, enabling a super-resolved map of their positions to be compiled.

We simulate a scene consisting of an array of fluorophores arranged in a spiral configuration (Fig.~\ref{Fig:supRes}(a)). The spacing of these fluorophores is below the diffraction limit, so the array is not resolved when imaged directly (Fig.~\ref{Fig:supRes}(b)). However, recording of multiple images of isolated fluorophores (an example shown in Fig.~\ref{Fig:supRes}(c))  yields a super-resolved image of the scene (Fig.~\ref{Fig:supRes}(d)). SI Movie 2 depicts this process. In this case the mean absolute error in position of a fluorophore is 129\,nm, a factor of 24 below the diffraction limit of $\lambda/(2\mathrm{NA})$ (where here $\lambda=633$\,nm, and $\mathrm{NA} = 0.1$). This is a promising result, which is readily extendable to 3D localisation microscopy~\cite{huang2008three}, and offers the exciting possibility of imaging with molecular specificity at unprecedented resolution deep inside living tissue. Of course in practice, the resolution of such a technique will depend critically on the fidelity of the inversion, the stability of the optical system and the signal-to-noise ratio in captured images.\\

\noindent{\bf Spectral bandwidth}\\
A key characteristic of our proposed fibre-inverter imaging system is the spectral bandwidth over which light can be unscrambled. The bandwidth of the TM of an MMF, $\Delta\lambda_{\mathrm{fib}}$, defines the range of wavelengths over which the same TM is valid~\cite{stellinga2021time}. Assuming that chromatic dispersion is negligible compared to spatial mode dispersion, $\Delta\lambda_{\mathrm{fib}}$ can be derived by considering the optical path length difference between a ray travelling at the critical angle of total internal reflection versus a ray travelling `straight through', parallel to the central fibre axis, yielding 
\begin{equation}\label{Eqn:bw}
    \Delta\lambda_{\mathrm{fib}}\sim2n_c\lambda^2/(L\mathrm{NA}^2),
\end{equation}
where $n_c$ is the refractive index of the core (see SI \S5 for derivation). This ray picture is equivalent to analysing the rate at which the PIMs with the highest and lowest phase velocities de-phase via propagation through the fibre. When the fibre is coupled to an inverter, we quantify the spectral bandwidth of the combined system by observing the rate of reduction in the mean power-ratio of imaged spots as the input wavelength is detuned from the design wavelength. See SI \S6 for more detail.

\begin{figure}[t]
   \includegraphics[width=\columnwidth]{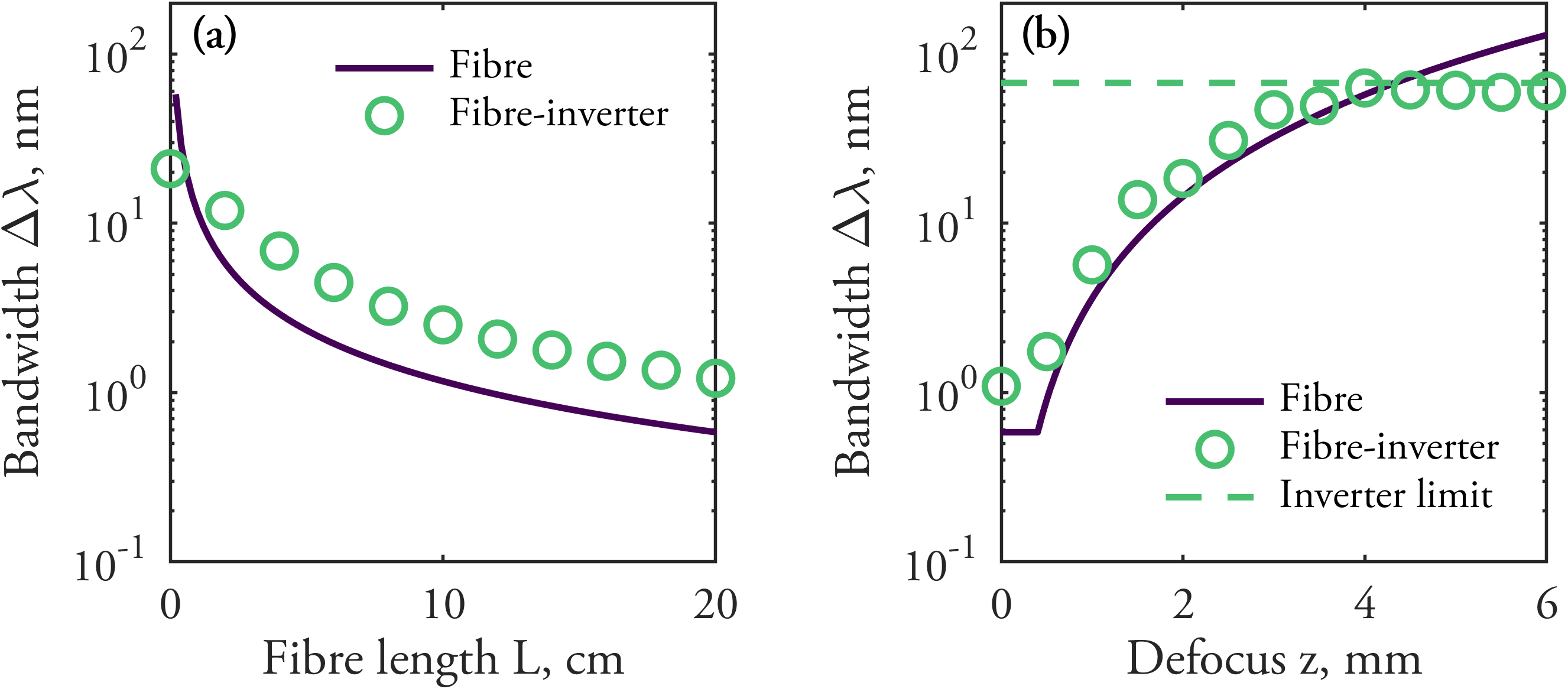}
   \caption{{\bf Spectral bandwidth}. ({\bf a}) Spectral bandwidth as a function of fibre length $L$ for our uncoupled fibre (purple line) and the combined fibre-inverter system (green circles), when the image plane is on the distal facet of the fibre. ({\bf b}) Spectral bandwidth for our uncoupled fibre (purple line) and the combined fibre-inverter system (green circles) as a function of image plane defocus depth $z$. Green dashed line indicates the inverter far-field bandwidth limit.}
   \label{Fig:bandwidth}
\end{figure}

\begin{figure}[t]
   \includegraphics[width=\columnwidth]{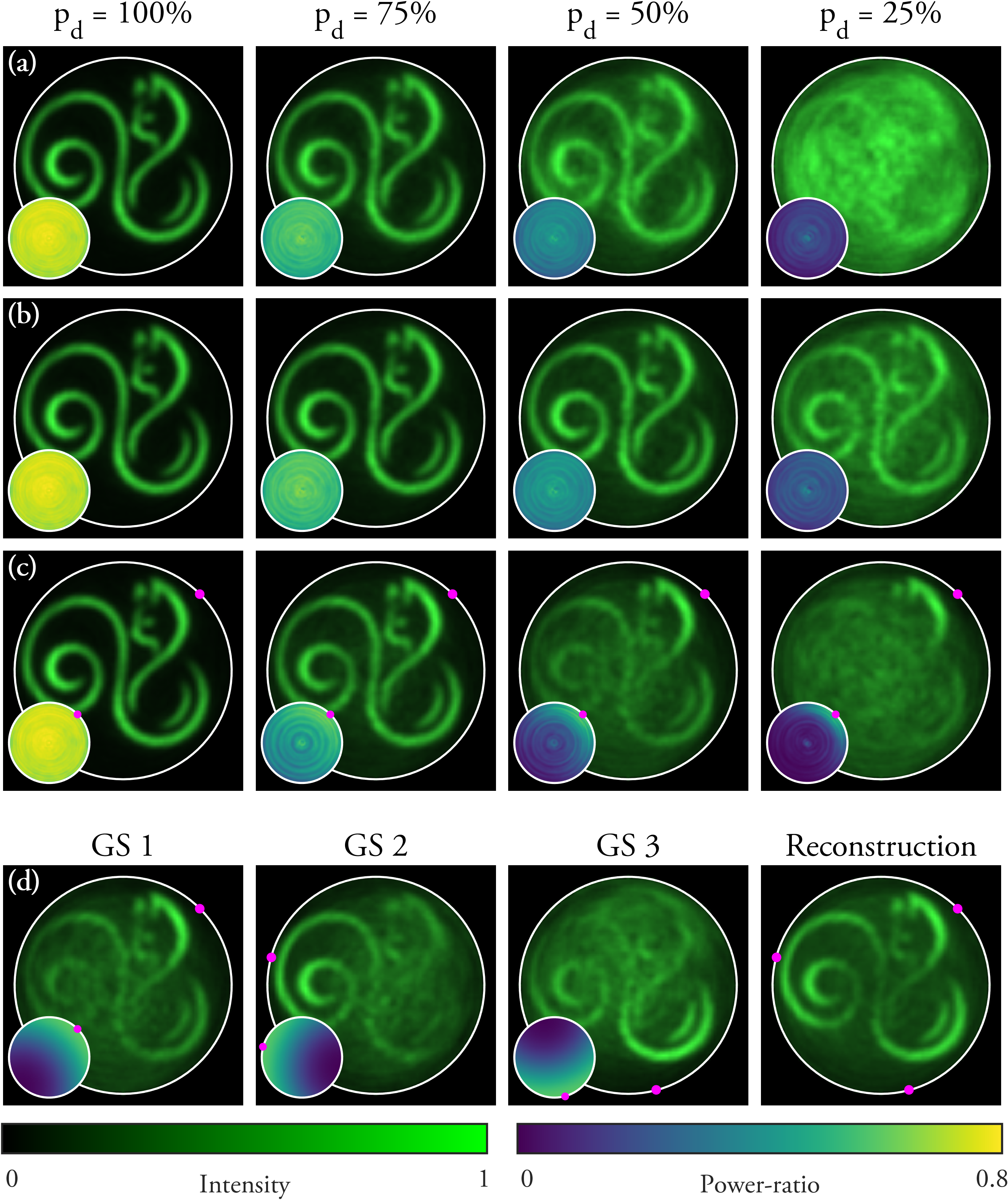}
   \caption{{\bf Inversion of imperfectly matched MMFs}. ({\bf a}-{\bf c}) Reduction in image quality as $p_{\mathrm{d}}$, the fraction of power on the diagonal of $\VecT'_{\mathrm{fib}}$, decreases. Insets show the power-ratio map in each case. ({\bf a}) In this case the inverter is non-adaptive and the correction phases are not modified. ({\bf b}) An improvement in imaging quality is obtained by adapting the correction phases assuming the full matrix $\VecT'_{\mathrm{fib}}$ is known. ({\bf c}) Even if $\VecT'_{\mathrm{fib}}$ is not known, we can open an isoplanatic patch around a guide-star (pink dot) placed at the distal facet. The size of this patch shrinks as $p_{\mathrm{d}}$ reduces. ({\bf d}) Multiple images recorded around 3 different guide-stars (GS 1-3, left 3 panels) can be combined (right most panel) to increase the field-of view. In this example $p_{\mathrm{d}} = 50\%$, and the reconstruction uses the estimated power-ratio maps shown inset in a weighted average of each image.}
   \label{Fig:imperfect}
\end{figure}

Figure~\ref{Fig:bandwidth}(a) shows the analytically calculated bandwidth of the fibre itself (purple line, given by Eqn.~\ref{Eqn:bw}), and the numerically simulated bandwidth of the combined MMF-inverter system (green circles) as a function of fibre length $L$.  Both curves follow a similar trend. The fibre itself limits the spectral bandwidth with which objects at the distal facet can be imaged, to about 1\,nm in the case of a 20\,cm long fibre.

Despite this narrow bandwidth, there is some tunability in the spectral response of the fibre-inverter system. We find that the wavelength of light that is brought into focus at the output plane can be tuned by adjusting the PIM-dependent phase delays imparted by the correction plane.
Irrespective of the fibre length, the spectral bandwidth over which it is possible to tune the wavelength is determined by the baseline bandwidth of the uncoupled inverter (i.e.\ as if paired with a fibre of zero length) $\Delta\lambda_{\mathrm{inv}}(L=0)\sim23$\,nm. This means a form of hyperspectral imaging may be possible: by synchronising the modulation of the corrector plane with a tunable spectral bandpass filter, images across a range of wavelengths could be recorded. See SI \S7 for a more in-depth discussion of this possibility.

Furthermore, when the image plane is moved away from the distal facet, the NA of rays emitted from a point on the image plane that are collected by the fibre is reduced below the NA of the fibre. Light emanating from adjacent points in the scene does not interfere in incoherent imaging, and so the reduction in NA for individual points has the effect of increasing the spectral bandwidth of the system. Therefore the bandwidth becomes a function of the image plane defocus distance $z$ according to 
\begin{equation}\label{Eqn:BWz}
  \Delta\lambda_{\mathrm{fib}}(z)\sim\begin{cases}
    (2n_c\lambda^2)/(L\mathrm{NA}^2), & \text{for}\,\, z\leq a/\mathrm{NA}\\
    (2n_c\lambda^2z^2)/(La^2), & \text{for}\,\, z>a/\mathrm{NA}.
  \end{cases}
\end{equation}
See SI \S5 for derivation and more detail on under which assumptions Eqn.~\ref{Eqn:BWz} holds. Figure~\ref{Fig:bandwidth}(b) shows how the spectral bandwidth of the fibre itself (purple line, given by Eqn.~\ref{Eqn:BWz}) and the combined MMF-inverter system (green circles, simulated) increase as the depth $z$ to the scene increases. For depths of $z\lesssim4$\,mm, the fibre itself limits the bandwidth. For greater levels of defocus, we find the inverter limits the spectral bandwidth to $\sim67$\,nm, which holds all the way to the far-field (green dashed line).\\

\noindent{\bf Adapting to imperfect MMFs}\\
Our MMF optical inversion strategy relies on the TM of the MMF being strongly diagonal when represented in the PIM basis. How realistic is this situation likely to be in practice? Power found in off-diagonal elements of the fibre's TM signifies mode coupling -- indicating that when the MMF is excited with an individual input PIM, other PIMs are excited upon reaching the output~\cite{li2021compressively}. Plöschner et al.\ have shown that with careful measurement, the TM of short lengths of step-index MMF can hold over 95\% of the power on the main diagonal when represented in the PIM basis -- i.e.\ the mode coupling is minimal~\cite{ploschner2015seeing}. Graded index MMFs are less predictable~\cite{flaes2018robustness}, hence here we have focussed on step-index MMFs, which at present show a greater prospect of being experimentally invertible.

Nonetheless, fibres exhibiting some degree of mode coupling are likely to be encountered in real-world scenarios. To study this we modify our model to incorporate MMF TMs that have varying levels of mode coupling and mode dependent loss. We introduce the imperfect fibre TM ${\VecT'_{\mathrm{fib}} \in \mathbb{C}^{P \times P}}$, and quantify the fraction of power on the main diagonal by $p_{\mathrm{d}}$. $\VecT'_{\mathrm{fib}}$ is constructed so that power is spread into off-diagonal elements in an experimentally realistic way, as detailed in ref.~\cite{li2021compressively}.

Figure~\ref{Fig:imperfect} shows the performance of our optical inverter when coupled with imperfect MMFs. Evidently, as $p_{\mathrm{d}}$ decreases, the inverter will less successfully unscramble the light emerging from the fibre. Fig.~\ref{Fig:imperfect}(a) shows the reduction in image quality when a fixed inverter -- designed for an ideal fibre -- is coupled to fibres possessing progressively increasing levels of modal coupling. Once ${p_{\mathrm{d}}\lesssim0.25}$ the image is no longer discernible. We find that mode coupling is more disruptive to imaging than mode dependent loss -- SI \S8 gives details.

We now consider how our inverter may adapt to improve image quality, and first analyse the case where the full TM of an imperfect MMF, $\VecT'_{\mathrm{fib}}$, has been measured.  One solution would be to redesign every phase mask in an attempt to map the device to a new inverse TM. However, this would mean that all 29 phase masks in our device would need to be re-configurable. With current technology, this is extremely challenging to achieve whilst also maintaining a high operating efficiency, as liquid-crystal SLMs are too lossy to be realistically used with this many (i.e.\ 29) reflections. Therefore we have aimed for a solution where only a single plane -- the correction plane -- need be reconfigurable. If $\VecT'_{\mathrm{fib}}$ is known, the modal phase delays imparted by the corrector plane can be adjusted and set to \begin{equation}\label{diagonalCorrection}
\boldsymbol\phi = -\arg\left[\mathrm{diag}\left(\VecP^{\dagger}\VecT'_{\mathrm{fib}}\VecP\right)\right],
\end{equation}
where $\phi_n$ is held in the $n^{\text{th}}$ element of column vector $\boldsymbol\phi$, and the operator $\mathrm{diag}(\cdot)$ creates a vector from the diagonal elements of a square matrix. In Eqn.~\ref{diagonalCorrection}, $\VecT'_{\mathrm{fib}}$ has been transformed to PIM-space, and the corrector plane phases are selected to undo the phase delays picked up by the uncoupled component of each PIM, while ignoring any mode coupling. Figure~\ref{Fig:imperfect}(b) shows image formation when adaptively adjusting the correction phases according to Eqn.~\ref{diagonalCorrection} as $p_{\mathrm{d}}$ reduces. We see the contrast is enhanced compared with the non-adaptive case -- thus imaging capabilities reduce gracefully rather than catastrophically failing upon the onset of mode coupling.\\

\noindent{\bf Imaging through flexible MMFs}\\
We have so far considered the pairing of our inverter with a static MMF whose TM has been accurately measured. However, the TM of an MMF is exceedingly sensitive to fibre configuration. Therefore if the fibre is free to move and bend, the TM is modified and is no longer accurately known. We now ask -- is there a way our inverter can accommodate this scenario?

The structure of our inverter has been designed to leverage fibre memory effects~\cite{li2021memory,li2021compressively,amitonova2015rotational}. Optical memory effects are associated with underlying correlations in a TM, that appear when it is represented in an appropriate basis~\cite{horstmeyer2015guidestar,li2021memory}. In the case of an MMF, its approximate cylindrical symmetry results in a rotational memory effect: rotating the input field causes a corresponding rotation in the output field~\cite{amitonova2015rotational}. The diagonal nature of the TM in the PIM basis also leads to a recently uncovered quasi-radial memory effect~\cite{li2021memory}. Our inverter gives us physical access to the optical fields in the PIM basis -- i.e.\ at the corrector plane -- enabling us to adapt the correction phases to take advantage of these memory effects. There are several ways we envisage these memory effects may be employed to invert flexible fibres:

Firstly, Plöschner et al.\ showed that if the changes in configuration of a short length of step-index MMF can be externally monitored, then the change to its TM can be {\it predicted} without needing to be re-measured~\cite{ploschner2015seeing}. This study showed that bending short lengths of fibre mainly altered the phase delay of individual PIMs, without introducing significant levels of modal cross-talk -- meaning that the inverse TM of curved MMFs can be well-captured by our inverter simply by updating the phase-delays given to each PIM at the corrector plane with those predicted from fibre curvature.

Secondly, Li et al.\ recently showed that an estimate of the TM of an MMF can be obtained without optical access to the distal end by deploying one or more fluorescent particles which can act as `guide-stars' located on the distal fibre facet~\cite{li2021memory}. In this case, the field at the proximal fibre facet that focuses onto a distal guide-star can be found by optimising the return fluorescent signal intensity -- in principle in just a few hundred milliseconds~\cite{horstmeyer2015guidestar}.
Knowledge of this proximal field, which here we denote by column vector ${\Vecu_{\mathrm{gs}} \in \mathbb{C}^{P}}$ when represented in the real-space basis, reveals the corrector plane phase delays $\boldsymbol\phi$ that bring this point into focus at the output of the inverter:
\begin{equation}\label{Eqn:guideStar}
    \boldsymbol\phi = -\left[\arg\left(\VecP^{\dagger}\Vecu_{\mathrm{gs}}\right)- \arg\left(\VecP^{\dagger}\Vect\right)\right],
\end{equation}
where column vector $\Vect \in \mathbb{R}^{P}$ specifies the location of the guide-star -- the element of $\Vect$ corresponding to the guide-star position is set to 1, and all other elements are set to 0.
The intuition behind Eqn.~\ref{Eqn:guideStar} is as follows: the first term on the right-hand-side provides an estimate of the relative phase of the PIMs on the target location at the distal end of the fibre. The second term captures the relative phase of the PIMs at the same lateral location at the {\it proximal} end of the fibre. Subtracting these values ensures that the calculated phase corrections only account for phase delays caused by propagation through the MMF itself.

Once light emanating from a guide-star at the distal facet of the MMF has been brought into focus at the inverter output, the existence of the rotational and quasi-radial memory effects mean other nearby points are simultaneously corrected. Therefore it is possible to image clearly through an area around the target point, known as an isoplanatic patch~\cite{li2021memory}. Figure~\ref{Fig:imperfect}(c) demonstrates image formation using this approach. The size and shape of the isoplanatic patch is governed by the magnitude of $p_{\mathrm{d}}$: if $p_{\mathrm{d}}=1$, the fibre TM is diagonal, and the isoplanatic patch extends across the entire distal fibre facet (Fig.~\ref{Fig:imperfect}(c), left-most panel). As mode coupling increases and $p_{\mathrm{d}}$ falls below 1, the isoplanatic patch contracts around the location of the guide-star -- see Fig.~\ref{Fig:imperfect}(c), and SI movie 3.

Evidently if the fibre exhibits significant mode coupling, then multiple guide-stars are required to scan a contracted isoplanatic patch across the whole distal facet. In this case, during calibration the light emanating from several guide-stars simultaneously could be distinguished via emission spectrum or speckle contrast~\cite{boniface2019noninvasive,boniface2020non}. Figure~\ref{Fig:imperfect}(d) shows a series of images as the isoplanatic patch is scanned over 3 separate guide-stars (left 3 panels), and a combined image formed from their weighted average (right hand panel). SI \S9 gives more details. While imaging in this way is no longer single-shot, many spatial modes may be unscrambled in parallel within each isoplanatic patch. This leads to an enhanced frame-rate compared with conventional wavefront shaping~\cite{vcivzmar2012exploiting}, with the enhancement dependent upon the level of mode coupling in the MMF. We highlight that this approach can also be used to enhance image contrast without the need for physical guide-stars if $\VecT'_{\mathrm{fib}}$ is known. In this case, the centre of the isoplanatic patch can be arbitrarily relocated (to a position specified by $\Vect$) using Eqn.~\ref{Eqn:guideStar} with ${\Vecu_{\mathrm{gs}} = \VecT'_{\mathrm{fib}}\Vect}$. We note there have also been several alternative proposals to reconstruct the TM of an MMF without access to the distal facet, which could potentially be used in conjunction with our inverter~\cite{gu2015design,gu2018improved,gordon2019characterizing}.

Lastly, it may be possible to engineer new types of MMF that are less sensitive to bending -- so that their TM does not alter significantly when their configuration changes~\cite{flaes2018robustness}.
Although such MMFs have not yet been manufactured, they represent a highly promising prospect if paired with our inverter in the future -- enabling high-speed incoherent imaging through flexible MMFs without the need for any adaptive feedback.\\

\noindent{\bf Discussion}\\
We now consider our proposal in the context of other approaches that aim to achieve single-shot imaging through MMFs. Kim et al.\ have demonstrated single-shot computational fluorescence microscopy through rigid MMFs~\cite{kim2014ultra,kim2015cannula}. In this approach, images of distal fluorescent objects are computationally reconstructed directly from the low contrast speckle patterns emanating from the proximal end (such as those shown in Fig.~\ref{Fig:defocus}(b-c), second column, for example). Calibration is first required to measure the incoherent intensity TM of the fibre or train a neural network~\cite{guo2022scan}. The main advantage of this approach is in its simplicity, as no additional optics are required at either end of the fibre during imaging. There are also no bandwdith limitations on the detected fluorescent signal. However, there is no opportunity to achieve super-resolution imaging with this method, and image reconstruction involves solving a highly ill-posed inverse problem which is very susceptible to measurement noise. In comparison, our concept essentially replaces the ill-posed computational image reconstruction step with all-optical image processing.

Machine learning based approaches have also been applied to demonstrate single-shot {\it spatially coherent} image transmission through MMFs~\cite{borhani2018learning}. These systems capture intensity images of high contrast spatially coherent speckle patterns emerging from the proximal end of a fibre, which are then processed by a pre-trained electronic neural network to reconstruct the intensity pattern of the corresponding field at the distal end. However, these systems rely upon constrained training data sets created by projecting patterns of light through fibres using spatially coherent laser sources. Such methods can only reconstruct images of artificial scenes formed from the same spatially coherent laser light at the distal facet. This currently precludes their application to image scenes emitting incoherent light, needed for the majority of real-world applications such as fluorescent micro-endoscopy~\cite{turtaev2018high,ohayon2018minimally}.

In the context of machine learning, our optical inverter can be interpreted as an all-optical diffractive deep neural network~\cite{lin2018all,luo2021computational} -- passively performing computational image reconstruction in a few nanoseconds. In this picture, the cascade of diffractive elements mimics the layered architecture of a linear convolutional neural network. For example, adjusting the phase delay imparted by a pixel on one phase plane modifies the diffracted field arriving at the next plane. This is analogous to tuning the weights of the coupling matrix linking the neurons between layers. In our case a linear network (i.e.\ without any non-linear activations) perfectly maps onto the linear inverse fibre TM.
The network `training' is carried out using fibre modes, which are the optimum basis to represent any possible image transmitted through the MMF. By dividing our diffractive network into three distinct subsections -- a mode sorting module, a phase correcting module and a mode combining module -- we impose a physically inspired model on the network's structure, in contrast to blindly optimising it.
Intriguingly, our modelling suggests that by physically realising an all-optical diffractive neural network, it is able to outperform its electronic counterpart for light scattering problems. In particular, our design can potentially unscramble {\it incoherent} optical signals in a single-shot manner with minimal loss of spatial resolution. We have also shown how super-resolution imaging may be possible without resorting to use of priors about the nature of the scenes. The root of this enhanced performance stems from the preservation of the phase and spatial coherence information emanating from the scene. In contrast, most electronic neural networks tasked with unscrambling light rely on training images capturing only the intensity of optical fields.

Finally, as our optical inverter is a passive element, no measurements of photon state are made until the image is detected at its output. Therefore our system may also preserve the quantum entanglement of photons sent through it, circumventing the need for active entanglement recovery~\cite{valencia2020unscrambling}, offering new opportunities in future quantum communication systems or quantum computing schemes.\\

\noindent{\bf Conclusion}\\
We have sketched out a new concept -- an `optical inverter' -- to simultaneously unscramble all light modes that have propagated through an MMF.
Our inverter can be understood as a bespoke scattering medium, designed to be complementary to an MMF so as to undo its optical effects. Although our work does not invoke transformation optics~\cite{mccall2018roadmap}, the action of the inverter is similar to `cloaking' an MMF within its spectral bandwidth, so that it is possible to see directly through it with no other computational steps necessary in the image reconstruction.

The key advantage of our concept is that it renders possible any form of wide-field microscopy at the tip of a hair thin strand of MMF. This includes localisation-based super-resolution imaging~\cite{betzig2006imaging,rust2006sub}, along with other emerging forms of parallelised super-resolution microscopy~\cite{bender2021circumventing}, and structured illumination microscopy~\cite{gustafsson2005nonlinear}.
Single-shot wide-field imaging at any distance beyond the distal end of a short length of MMF also becomes possible. These new imaging modalities come at the cost of a relatively complicated design and currently with a limited spectral bandwidth -- at least when imaging the distal fibre facet. However, the reliance of our concept on multi-plane light conversion means it is within the reach of current fabrication technology, and future improvements in design may yield broader bandwidth operation. Furthermore, by introducing only a single re-configurable phase mask, we have shown how our optical inverter can adapt to a range of fibre inverse TMs: a feature that also represents the first example of a new class of MPLC device structured to actively incorporate optical memory effects.

The optical inversion strategy we have described here can potentially be extended to unscramble light that has passed through other objects, such as photonic crystal waveguides~\cite{birks1997endlessly}, photonic lanterns~\cite{choudhury2020computational} or biological tissue~\cite{boniface2020non}. Finally, we anticipate that all-optical inversion of scattered light will find an array of applications beyond optical imaging: benefiting the fields of mode division multiplexing for high capacity optical communications~\cite{kahn2017communications,zhou2021high}, as well as quantum cryptography~\cite{bechmann2000quantum} and classical and quantum optical computing~\cite{carolan2015universal,leedumrongwatthanakun2020programmable}.

\section{Acknowledgements}
We thank Dr Stephen Simpson for useful discussions. DBP thanks Dr Joel Carpenter for useful advice on MPLC design. DBP thanks the Royal Academy of Engineering, and the European Research Council (Grant no.\ 804626) for financial support. TC acknowledges the European Research Council for financial support (Grant no.\ 724530).

\section{Contributions}
DBP and TC developed the idea for the project. DBP conceived the methods of adaptive MMF inversion and supervised the project. UGB and HK optimised the fibre mode sorters. UGB performed all imaging simulations and spectral analysis, with support from DBP and HK. DBP, UGB and HK wrote the paper, with editorial input from TC.

\bibliography{refs}

\onecolumngrid
\newpage

\setcounter{equation}{0}
\setcounter{figure}{0}
\vspace{50cm}
\noindent{\centering{\LARGE Supplementary Information}\par}
\vspace{5mm}

\begin{enumerate}
\item Fibre mode sorter design and performance quantification.
\item Numerical modelling of the MMF-inverter system.
\item Description and schematics of envisaged experimental set-up.
\item MMF-inverter power ratio map.
\item Theoretical spectral bandwidth of a MMF.
\item Numerical evaluation of the spectral bandwidth of the MMF-inverter system.
\item Effect of mode dependent loss.
\item Adapting the inverter for hyperspectral imaging.
\item Combining images from different isoplanatic patches.
\item Description of Supplementary movies.
\end{enumerate}

\newpage

\section{\S1: Fibre mode sorter design and performance quantification}
A multi-plane light converter (MPLC) consists of a series of 2D diffractive planes separated by free-space. These planes each impart a spatially varying phase delay to transmitted light. The phase delay profiles are optimised in order to attempt to convert one set of input fields to another set of output fields. In this work these sets are the propagation invariant fibre modes (as inputs) and the spatially separated Gaussian spots (as outputs). Here phase planes are designed using an optimisation procedure known as the wavefront matching method \cite{hashimoto2005optical}, which is similar to a gradient descent optimisation. A detailed description of this algorithm is given in ref.~\cite{fontaine2019laguerre}. Here we provide a brief descriptive overview: to start with, the input modes are individually numerically propagated from the input plane, through all the phase masks up to the last one. The output modes are individually back-propagated from the output to the opposite side of the last plane, where they meet the input modes. During this initialisation step we set the phase masks to have a profile of zero relative phase everywhere; this does not have to be the case – any other phase pattern can be used (which may then converge to a different local solution). At the last plane, each input mode is `overlapped' with its corresponding output mode. The overlap operation is a Hadamard product (i.e.\ element by element multiplication) between each input mode and the conjugate of the corresponding output mode. All of the overlaps are then averaged, and the argument is taken. The resulting phase profile is then added to the current phase profile of the mask between where the modes have met, generating a new mask which is used in the following iterations. This processes is repeated to optimise each phase mask in turn, and then the optimisation of all masks is cycled for a number of iterations. Over consecutive iterations the phases profiles converge to a solution. Depending on the particular mode sets and mode number the algorithm can take tens to hundreds of iterations to converge. During the design process, and subsequent modelling of imaging performance, we assume each plane is infinitely thin and that there is no cross coupling within each plane. These assumptions have recently worked well to design experimentally realised MPLCs~\cite{fontaine2019laguerre,mounaix2020time}.

The physical dimensions of the masks in our design are 0.5x0.5\,cm with a pixel pitch of \SI{12.5}{\micro\meter}, giving a mask resolution of $400\times400$\, pixels. The masks are separated by 3.2\,cm. These dimensions are similar to MPLCs that have been successfully realised experimentally. Note also the image of the proximal end of the fibre has been magnified $\sim$10 times to a radius of \SI{426}{\micro\meter}, which is a suitable size to make sure that no light spills off the edge of the masks.

\begin{figure}[H]
\centering
\includegraphics[width=16cm]{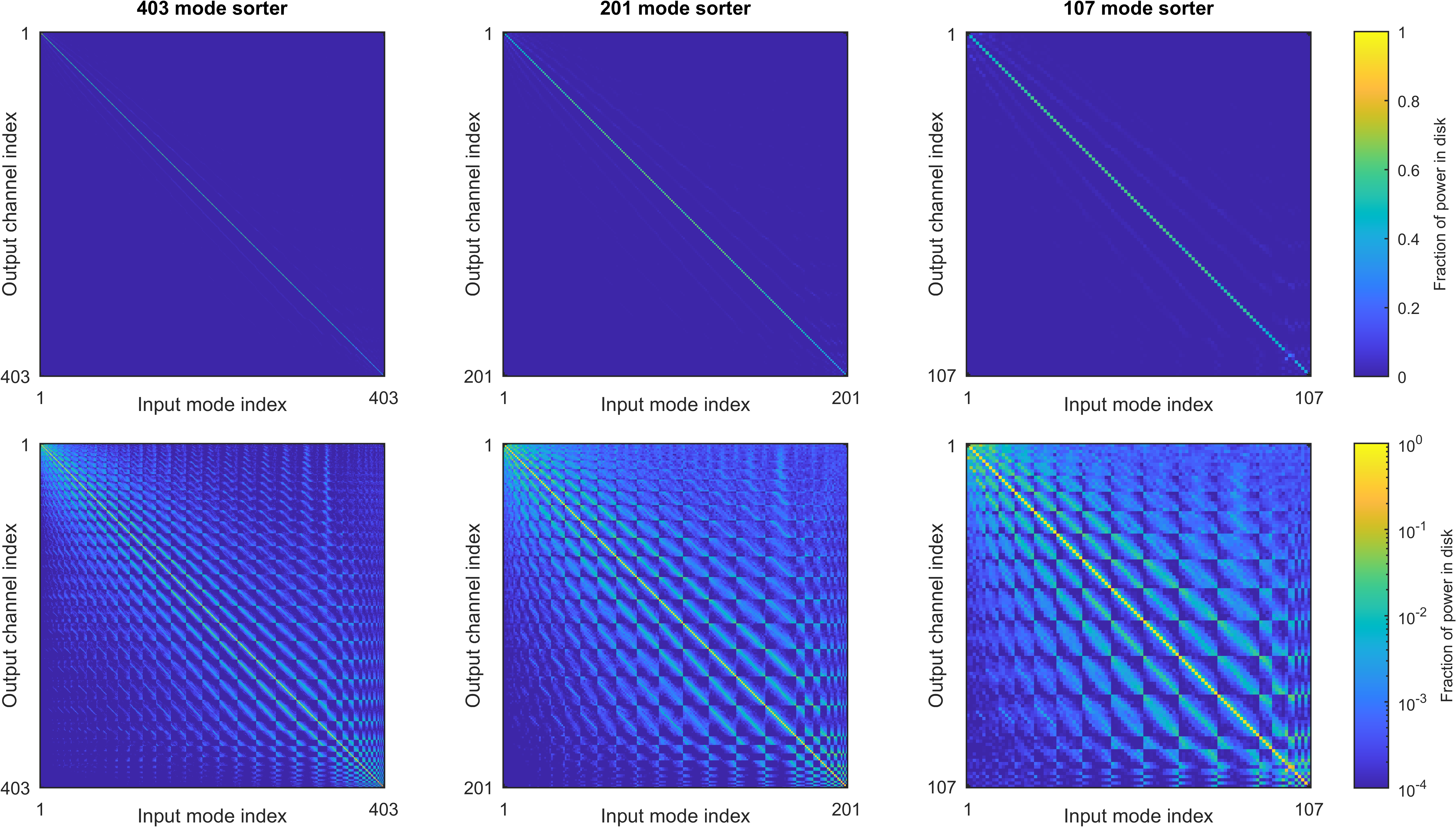}
\caption{{\bf Modal cross-talk}. Coupling matrices of the 403-mode 15-plane sorter discussed in the main paper (left), a 201-mode 11-plane sorter (middle), and a 107-mode 7-plane sorter (right). The matrices are plotted on a linear scale in the first row, and on a logarithmic scale in the second row to reveal detail.}
\label{fig:crosstalk}
\end{figure}
Once designed, the performance of the sorter is characterised by its efficiency and cross-talk, which are defined as follows. We define an output channel – a circular area as indicated in Fig.~2(d) in the main text – for every Gaussian spot in the output plane. For every input mode we then calculate the fraction of the total power in every output channel - these values are arranged into a column vector. Retrieving the same information for all input modes forms a coupling matrix, as shown in Fig.~\ref{fig:crosstalk}. The diagonal elements of this matrix indicate the efficiency with which every mode is sorted into its designated channel, and the off-diagonal elements show the amount of signal contamination when light from one channel bleeds into its neighbours. The cross-talk of every mode is given by the ratio of the fractional power in the designated output channel to the fractional power in all other output channels. For an ideal sorter the coupling matrix would be an identity matrix. More formally, the efficiency $e_n$ and cross-talk $c_n$ of the $n$-th mode are defined as:
\begin{equation}
    e_n = \frac{\displaystyle\iint \left|\phi_n(\vec{r}) \right|^2 dS_n(\vec{r})}{\displaystyle\iint \left|\phi_n(\vec{r}) \right|^2 dS(\vec{r})}; \quad\quad\quad\quad\quad
    c_n = \frac{\displaystyle\iint \left|\phi_n(\vec{r}) \right|^2 dS_n(\vec{r})}{\displaystyle\sum_{i=1}^{N}\displaystyle\iint \left|\phi_n(\vec{r}) \right|^2 dS_i(\vec{r})},
\end{equation}
where $\phi_n$ is the complex field of the $n$-th output mode, $S_n$ is the output channel area of the $n$-th mode, $S$ is the area of the entire field-of-view (i.e.\ the area of the image of the fibre core), and $\vec{r}$ is the coordinate vector.

We optimised several MPLC devices able to sort different numbers of modes using different numbers of planes. The efficiency and cross-talk of three such sorters is shown in Fig.~\ref{fig:crosstalk}, which have design criteria and average efficiency and cross-talk given in table~\ref{tb:expResults}:
\begin{table}[H]
\begin{center}\begin{tabular}{c||c|c|c|}
No. of modes & No. of planes & Efficiency, \% & Cross-talk, \%\\
\hline\hline
107 & 7 & 53.0 & 18.4\\
201 & 11 & 54.1 & 20.0\\
403 & 15 & 50.8 & 23.1
\end{tabular}\end{center}
\caption[]{\textbf{Sorter performance}}
\label{tb:expResults}
\end{table}
We highlight that the levels of efficiency and cross-talk are very promising, especially when considering the unusually high mode capacity of these MPLC designs.

\newpage

\section{\S2: Numerical modelling of the MMF-inverter system}
To model the performance of the fibre-inverter system, we first calculate the coherent TM of a length of ideal MMF, ${\VecT_{\mathrm{pix}} \in \mathbb{C}^{P \times P}}$, represented in real-space (i.e.\ in the $P-$dimensional pixel basis) by solving the wave-equation to find the circularly polarised PIMs and their associated phase velocities $\beta$. Thus
\begin{equation}\label{Eqn:fibreTMSI}
    \VecT_{\mathrm{pix}}=\VecP\VecD(L)\VecP^\dagger.
\end{equation}
Here $\VecP \in \mathbb{C}^{P \times N}$ is a matrix transforming from $N$-dimensional PIM-space to real-space. The $n^{\text{th}}$ column of $\VecP$ is calculated by representing the complex field of the $n^{\text{th}}$ PIM on a pixelated Cartesian grid, and reshaping this into a column vector. The chosen reshaping format is henceforth used consistently to map 2D pixelated real-space to a 1D representation. $\VecD(L) \in \mathbb{C}^{N \times N}$ is a diagonal unitary matrix. The $n^{\mathrm{th}}$ diagonal element of $\VecD$ holds the phase change $\exp({i\beta_n L})$ experienced by the $n^{\mathrm{th}}$ PIM after propagation through an MMF of length $L$.

To model the inverter we use Eqns.~\ref{Eqn:MPLCRealTM} and~\ref{Eqn:MPLC} in the main text. Here matrix $\VecH \in \mathbb{C}^{P \times P}$ is a free-space propagation matrix transforming the field from the output facet immediately after plane $m$ to the input facet immediately before plane $m+1$. It is found using the angular spectrum method: $\VecH = \mathcal{F}^{-1}\VecK(\delta z)\mathcal{F}$, where $\mathcal{F} \in \mathbb{C}^{P \times P}$ is the 2D Fourier transform matrix, and $\VecK(\delta z) \in \mathbb{C}^{P \times P}$ is unitary diagonal matrix capturing the phase change picked up by each Fourier component over propagation distance $\delta z$. Matrix $\VecC$ represents the TM of the corrector plane. It is a diagonal matrix holding the phase imparted by each pixel of the corrector plane on its diagonal. 

Matrix $\VecL$ transforms from the fibre output, through the cylindrical mode converter to the MPLC input plane. The mode converter performs optimally when the proximal end of the fibre is imaged directly in between the two cylindrical lenses. In our simulation $\VecL$ is given by
\begin{equation}
    \VecL = \VecH_\text{cyl2s} \VecA_\text{cyl} \VecH_\text{cyl} \VecA_\text{cyl} \VecH_\text{cyl,1/2}^*,
\end{equation}
where $\VecH_\text{cyl,1/2}^*$ is a backwards free-space propagator from the mid-point between the two cylindrical lenses, $\VecH_\text{cyl}$ is the free-space propagator from the first cylindrical lens to the second, $\VecH_\text{cyl2s}$ is the free-space propagator from the second cylindrical lens to the sorter, and $\VecA_\text{cyl}$ is a diagonal matrix capturing the phase mask of a cylindrical lens on its diagonal, i.e.\ the 2D phase mask function is given by $\exp\left( {-i\pi\frac{\lambda}{f_\text{cyl}}Y^2} \right)$. Here $f_\text{cyl}$ is the focal length of the lens, $Y$ is a 2D array with the y-coordinates of the phase mask, and we take the element-wise square. This 2D cylindrical phase function is reshaped into a column vector to form the diagonal elements of matrix $\VecA_\text{cyl}$.

It is important to note that under ideal circumstances the combiner would perform the inverse operation to that of the sorter, i.e.\ Eqn.~\ref{Eqn:MPLCRealTM} would contain conjugate transposes instead of transposes. In reality, when modes propagate through the combiner they pick up further phase delays which do not undo the phase delays picked up while propagating through the sorter. Our inverter design allows to eliminate these phase shifts at the corrector plane.

A note on polarisation: Our simulations are scalar in nature. Since the PIMs do not couple to the orthogonal circular polarisation channel during propagation along a short length of MMF, we model transmission of a single handedness of circular polarisation through the fibre. We essentially simulate the case where a polarisation filter passes only one circular polarisation onwards through the inverter. As the phase profiles of each plane of the inverter are smoothly varying, we anticipate minimal polarisation coupling of light propagating through the inverter.

A note on our choice of simulation method: Here we have described our MMF-inverter model in matrix form. However, the matrices describing the inverter are very high dimensional. Therefore, rather than constructing these very large matrices, in practice we simulate the propagation of light through the inverter one mode at a time using, for example, 2D fast Fourier transforms and phase additions. These operations can then be parallelised to some extent to speed up our simulations without requiring large amounts of computer RAM.

\newpage

\section{\S3: Description and schematics of envisaged experimental set-up}
\begin{figure}[H]
\centering
\includegraphics[width=18cm]{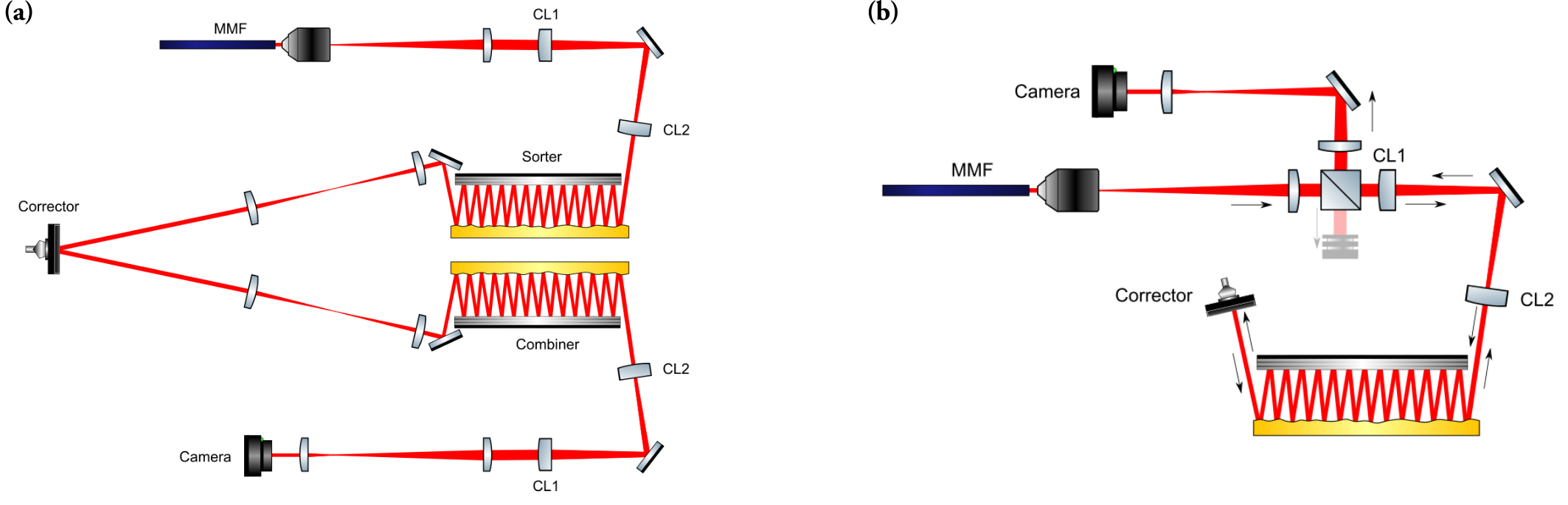}
\caption{{\bf Possible experimental implementations of an MMF optical inverter} ({\bf a}) A schematic showing independent mode sorting and mode combining MPLCs. ({\bf b}) Schematic of a more compact but less efficient set-up where the mode sorter and combiner are combined into one device. CL = cylindrical lens.}
\label{fig:opt_setup_combined}
\end{figure}

In the following we describe how our optical inverter may be experimentally realised in the future. We are currently working towards this goal. Figure~\ref{fig:opt_setup_combined}(a) shows a schematic of a possible optical system.

A scrambled optical field emerging from the proximal fibre facet is magnified by the combination of objective and conventional tube lenses. The fibre facet is re-imaged to a plane exactly in between the two following cylindrical lenses which form a cylindrical mode converter. The light next enters a reflective MPLC formed from a dielectric mirror opposite a gold coated lithographically etched diffracting mirror. The latter encodes the optimised phase profiles in its surface height. The planes are situated adjacently to one another along this mirror. The separation between the dielectric mirror and the contoured mirror is set to be equal to half the free-space propagation distance between the MPLC planes ($d$). Here we assume that the dominant angle of incidence and reflection is close to normal. After the 14$^{\mathrm{th}}$ reflection from the diffractive element, the field is re-imaged by a 4f system of two lenses to a plane a distance $d$ in front of the corrector plane SLM. The upper MPLC acts as a fibre mode sorter -- thus, the PIMs arrive spatialy separated at the corrector plane.

It is now possible to apply digitally tunable PIM-dependent phase corrections to these modal components by choosing the spatially varying phase pattern to display on the SLM. These phase corrections encompass three components: (i) the phase corrections required to undo propagation through the MMF, (ii) additional phase corrections to remove mode dependent phase delays generated by the mode sorter, and (iii) the phase profiles of the 15$^{\mathrm{th}}$ mode sorting MPLC plane and the 1$^{\mathrm{st}}$ mode combining MPLC plane which are imaged onto one another at the SLM.

The light then reflects from the SLM, and follows a symmetrical path back through a second MPLC that now forms the mode combiner. Finally, the light passes through a second cylindrical mode converter. In this way the PIMs are recombined and imaged onto a camera where the image of the field at the distal fibre facet is formed. Defocussing of the image plane away from the end of the fibre is achieved simply by moving the camera axially. 

Our optical setup can also be built in a more compact way, as shown in Fig.~\ref{fig:opt_setup_combined}(b). Now, the corrector SLM not only introduces phase delays to different parts of an incoming beam, but also acts as a mirror that steers light back to the same MPLC system. The direction of light propagation is shown with black arrows. This second configuration is simpler, but also less efficient as 50\% of the light is lost at each of the two passes through the non-polarising beam splitter.

\newpage

\section{\S4: MMF-inverter power ratio map}
\begin{figure}[H]
\centering
\includegraphics[width=8cm]{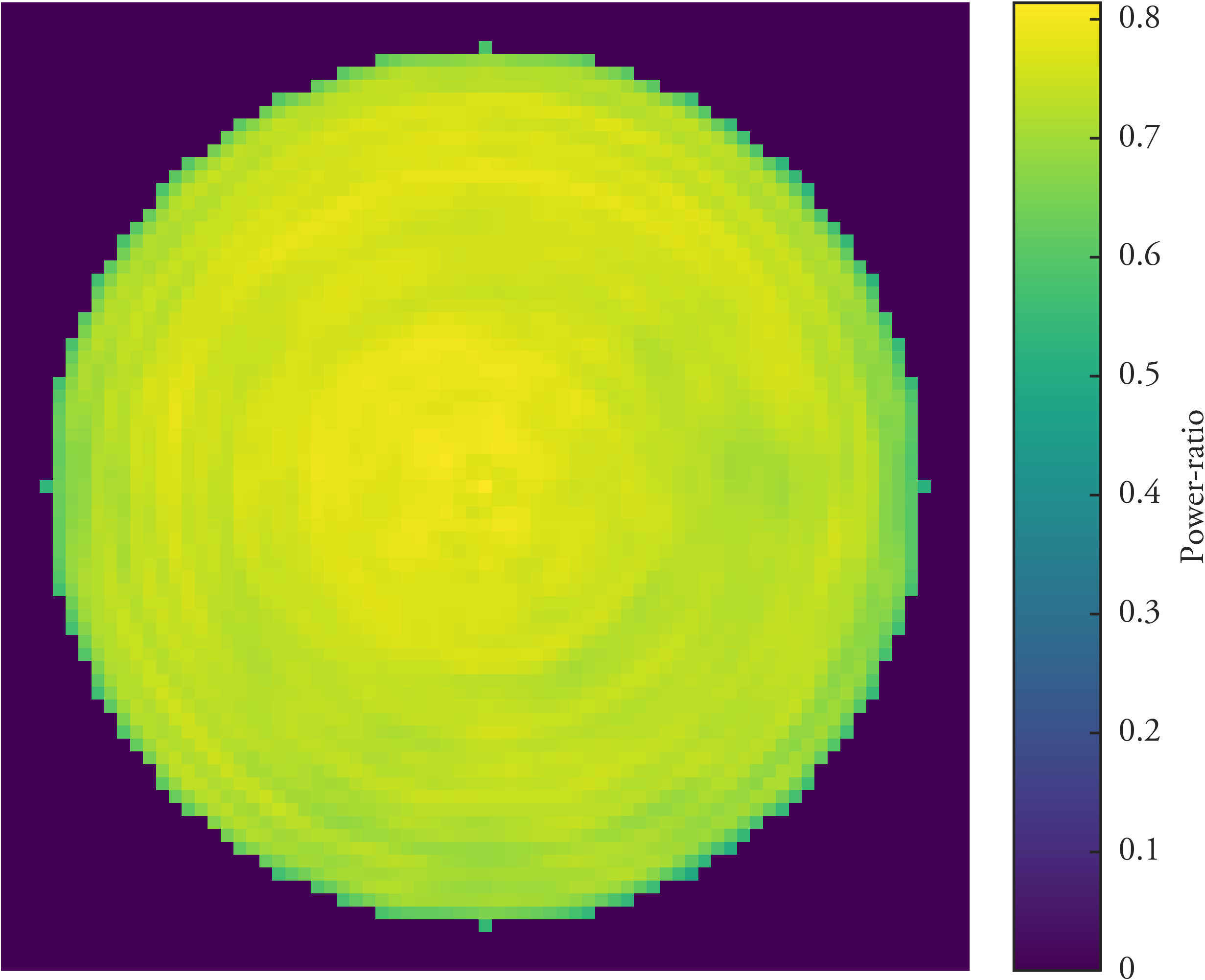}
\caption{{\bf Power-ratio map of our inverter coupled with an ideal fibre.}}
\label{fig:powMap_idealFibre}
\end{figure}

\vspace{1cm}
\section{\S5: Theoretical spectral bandwidth of a MMF}
Firstly we derive the bandwidth of a fibre when imaging on the distal facet. We make use of Snell's law to account for refraction of light entering the fibre core: $\sin{\theta} = n_c\sin{\phi} = \mathrm{NA}$. Here $\theta$ is the angle of critical ray outside fibre core. $\phi$ is the angle of refracted critical ray inside the fibre core. Therefore the critical angle inside the fibre is given by
\begin{equation}
    \phi = \sin^{-1}\left(\mathrm{NA}/n_c\right).
\end{equation}
We now define the optical path length difference  $\Delta_{\mathrm{OPL}}$ between the fastest and slowest rays travelling along a fibre of length $L$.
\begin{equation}\label{Eqn:opl}
    \Delta_{\mathrm{OPL}} = \frac{n_cL}{\cos{\phi}}-n_cL.
\end{equation}
We substitute for $\phi$ and make use of the identity ${\cos(\sin^{-1}{x}) = (1-x^2)^{\frac{1}{2}}}$ to give:
\begin{equation}
    \Delta_{\mathrm{OPL}} = n_cL\left[\left(1-\frac{\mathrm{NA}^2}{n_c^2}\right)^{-\frac{1}{2}} - 1\right].
\end{equation}
To further simplify this we Taylor expand the first term in the square brackets, ignoring terms of higher powers than $(\mathrm{NA}/n_c)^2$ (valid for typical low NA fibres of $\mathrm{NA}\lesssim0.22$):
\begin{equation}
    \left(1-\frac{\mathrm{NA}^2}{n_c^2}\right)^{-\frac{1}{2}}\sim 1 + \frac{\mathrm{NA}^2}{2n_c^2}.
\end{equation}
Therefore
\begin{equation}
    \Delta_{\mathrm{OPL}} = \frac{\mathrm{NA}^2L}{2n_c}.
\end{equation}
Now using ${\tau = \Delta_{\mathrm{OPL}}/c}$ (where $\tau$ is the time difference between light emerging from the fibre travelling the slowest and fastest paths), and ${\Delta f = 1/\tau}$ (where $\Delta f$ is the associated frequency bandwidth), and ${\Delta\lambda = \lambda^2\Delta f/c}$, we can find the bandwidth of the fibre TM:
\begin{equation}\label{Eqn:bandwidthL}
    \Delta\lambda = \frac{2\lambda^2n_c}{L\mathrm{NA}^2}.
\end{equation}
We note that we have ignored chromatic dispersion (i.e.\ any dependence of $n_c$ on $\lambda$), as we expect spatial mode dispersion to dominate over the wavelength range we consider.

Next we use a similar approach to calculate the TM bandwidth as a function of both fibre length ($L$) and the distance from the distal fibre facet to the scene ($z$). In this case, instead of considering the optical path length difference between the ray at the critical angle of total internal reflection and the ray that travels `straight through' the fibre (as above), we now consider the optical path length difference of the two extreme rays emanating from a single point in the scene that pass through the fibre - as it is only these rays that must interfere constructively at the far-field of the proximal fibre facet.  Let the angle between these rays leaving a point in the scene at the centre of the field-of-view be $\alpha$, and thus
\begin{equation}
    \tan{\alpha} = \frac{a}{z}.
\end{equation}
Again using Snell's law to find the refracted angle of the ray inside the fibre core we have
\begin{equation}
    \sin{\alpha} = n_c\sin{\phi} = \sin\left[\tan^{-1}\left(\frac{a}{z}\right)\right]
\end{equation}
Therefore $\phi$ is given by
\begin{equation}
    \phi = \sin^{-1}\left[\frac{a}{n_cz}\left(1+\frac{a^2}{z^2}\right)^{-\frac{1}{2}}\right],
\end{equation}
where here we have made use of the trigonometric identity ${\sin(\tan^{-1}x) = x(1+x^2)^{-\frac{1}{2}}}$. Substituting this into Eqn.~\ref{Eqn:opl} and again using the identity ${\cos(\sin^{-1}{x}) = (1-x^2)^{\frac{1}{2}}}$ we have
\begin{equation}
    \Delta_{\mathrm{OPL}} = n_cL\left(\gamma-1\right),
\end{equation}
where
\begin{equation}
    \gamma = \left(1-\frac{a^2}{n_c^2(z^2+a^2)}\right)^{-\frac{1}{2}}.
\end{equation}
Therefore
\begin{equation}\label{Eqn:bandwidthd}
    \Delta\lambda = \frac{\lambda^2}{n_cL}\left(\gamma-1\right)^{-1}.
\end{equation}
When $z>>a$, Eqn.~\ref{Eqn:bandwidthd} simplifies to
\begin{equation}\label{Eqn:bandwidthd2}
    \Delta\lambda = \frac{2n_cz^2\lambda^2}{La^2},
\end{equation}
which can be found by making the small angle approximation ${\tan{\alpha} \sim \sin{\alpha}}$ at the start of the derivation, which is equivalent to approximating $\gamma$ according to:
\begin{equation}
    \gamma\sim \left(1-\frac{a^2}{n_c^2z^2}\right)^{-\frac{1}{2}} \sim 1+\frac{a^2}{2n_c^2z^2}.
\end{equation}
A note on the limits:  in the far-field of the fibre facet in the limit $z\rightarrow\infty$, Eqn.~\ref{Eqn:bandwidthd} shows that $\Delta\lambda\rightarrow\infty$. In reality, diffraction effects (which are not included in our ray-approximation based derivation) and chromatic dispersion will limit the bandwidth. In addition, Eqn.~\ref{Eqn:bandwidthd} is valid only for the range $z>a/\mathrm{NA}$, as when $z$ is smaller than this limit, $\phi$ is equal to the critical angle of total internal reflection and $\Delta\lambda$ is independent of $z$ and given by Eqn.~\ref{Eqn:bandwidthL}.

\section{\S6: Numerical evaluation of spectral bandwidth of the MMF-inverter system}

\begin{figure}[H]
\centering
\includegraphics[width=14cm]{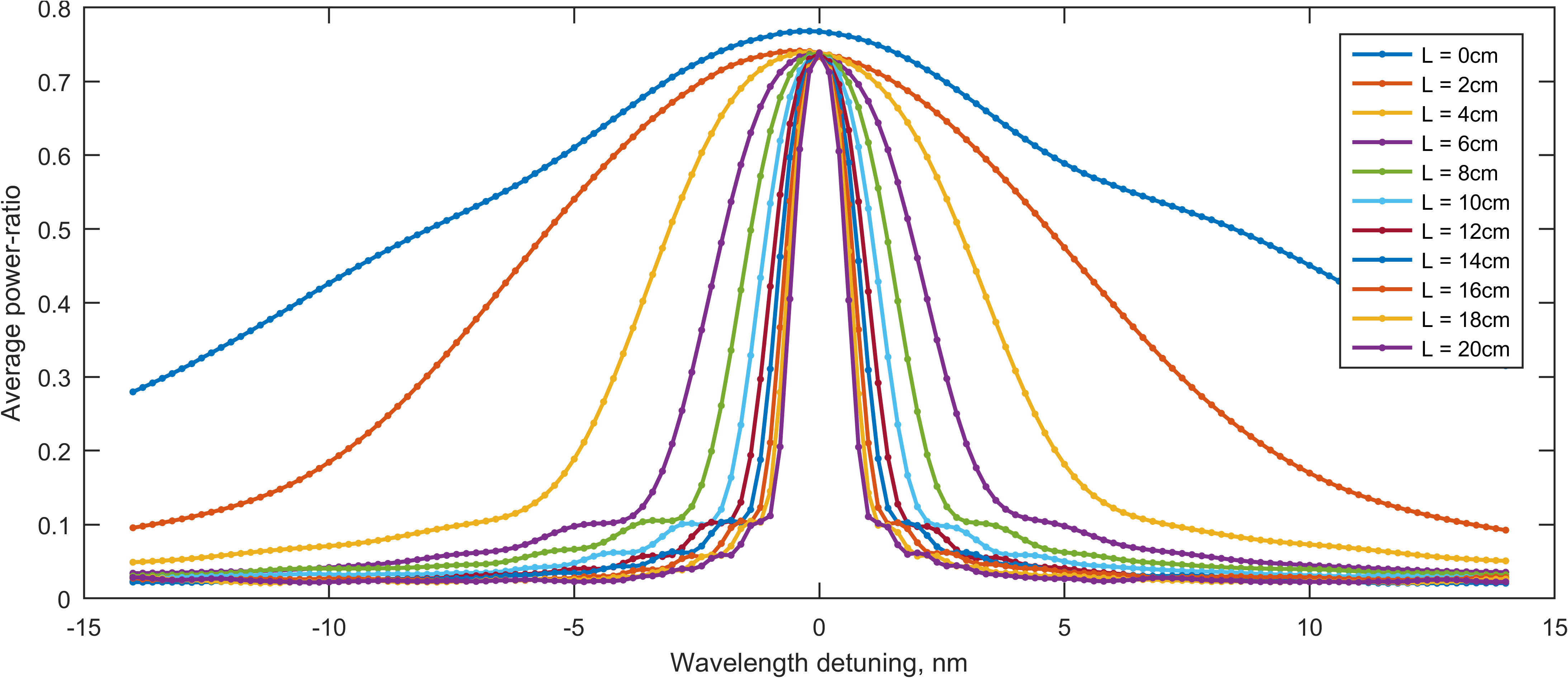}
\caption{\bf Spectral performance of the fibre-inverter system}
\label{fig:spectral_band}
\end{figure}
To quantify the spectral performance of the fibre-inverter system we look at how the average power-ratio changes with the wavelength. The power-ratio map, such as shown in SI Fig.~\ref{fig:powMap_idealFibre} is built up by propagating each pixel inside the fibre core through the fibre-inverter system and calculating the fraction of power that arrives into the diffraction limited spot corresponding to the input pixel. The average across all pixels is then taken. The same is repeated for detuned wavelengths to build up a distribution as shown in Fig.~\ref{fig:spectral_band}. Fitting a Gaussian to find the FWHM then gives the spectral bandwidth $\Delta\lambda$ as plotted in Fig.~4(a) in the main paper. When evaluating the bandwidth for planes defocused from the distal end of the fibre (for the data in Fig.~4(b) in the main paper) we do the same as above, but only for one central input pixel to reduce the significant computational cost required for this simulation.
When replacing the original wavelength $\lambda_0$ for which the phase masks of the inverter have been designed with a detuned wavelength $\lambda_{\mathrm{det}}$, we simulate how this will change the propagation through both the fibre and the inverter: In the case of the fibre, we solve the monochromatic wave equation to find a new set of PIMs and corresponding phase velocities at each new wavelength $\lambda_{\mathrm{det}}$, enabling construction of a new fibre propagation matrix $\VecT_{\text{fib}}(\lambda_{\mathrm{det}})$. We note also that this may change the number of modes supported by the fibre. In the case of the inverter, we update all the phase masks (including the corrector plane) to account for the change in imparted phase delay as a function of wavelength. This is achieved by multiplying the original phase-delay by the ratio of the design and detuned wavelengths $\lambda_0/\lambda_{\mathrm{det}}$~\cite{mitchell2017polarisation}. 

\newpage

\section{\S7: Adapting the inverter for hyperspectral imaging}
\begin{figure}[H]
\centering
\includegraphics[width=14cm]{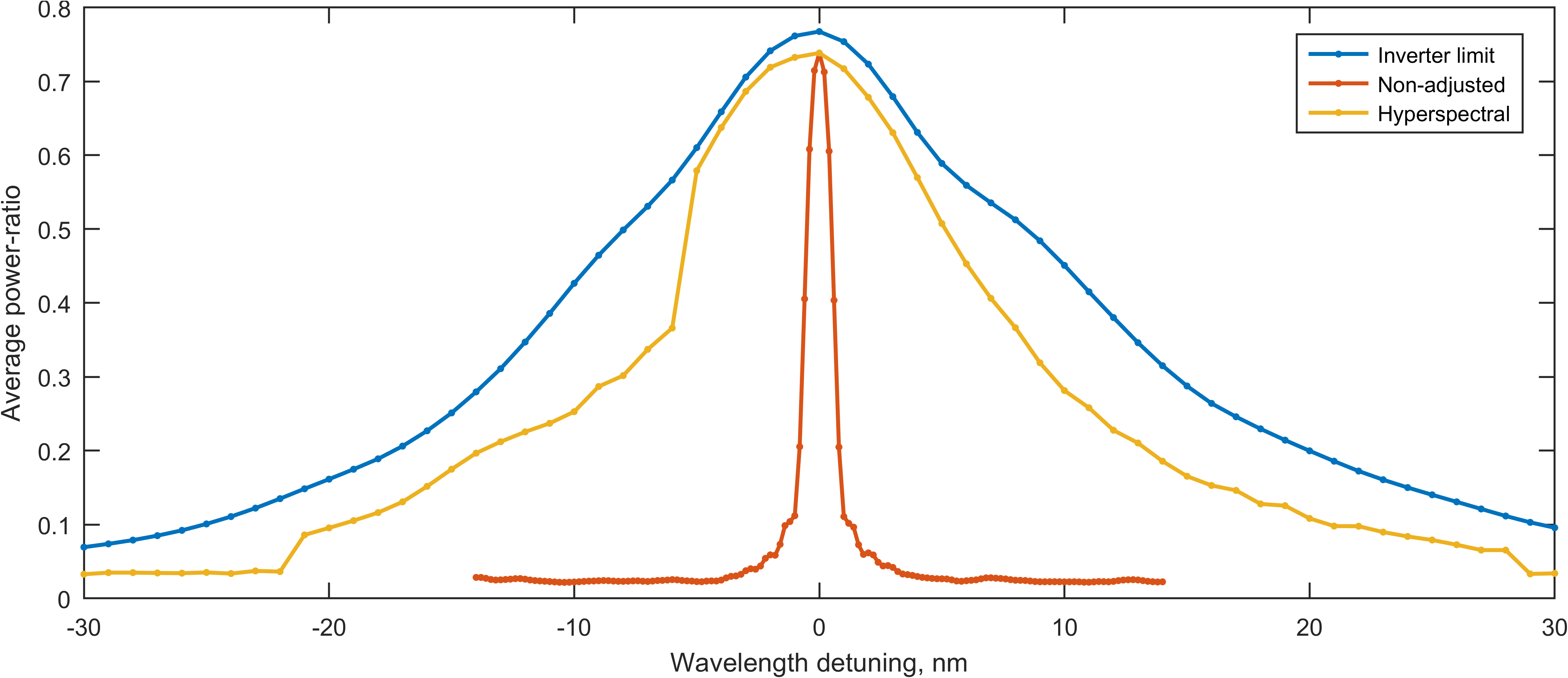}
\caption{\bf Hyperspectral performance of the fibre-inverter system}
\label{fig:spectral_band_hyper}
\end{figure}
Since the spectral bandwidth of our inverter (when not coupled to a fibre) is significantly larger than the bandwidth of the fibre itself, our system allows for the possibility of hyperspectral imaging. The phases on the corrector plane can be adjusted for different wavelengths which, together with a tunable spectral bandpass filter, would allow images across a range of wavelengths to be recorded. Shown in the figure below is the spectral response (obtained as described in the section above) of the inverter coupled with a 20cm long fibre, when the corrector phases are adapted to each detuned wavelength individually (yellow). Evidently the bandwidth far exceeds the non-adjusted case (orange), and approaches the bandwidth of the inverter itself (blue).

\newpage

\section{\S 8: Effect of mode dependent loss in the fibre itself}
\begin{figure}[H]
\centering
\includegraphics[width=16cm]{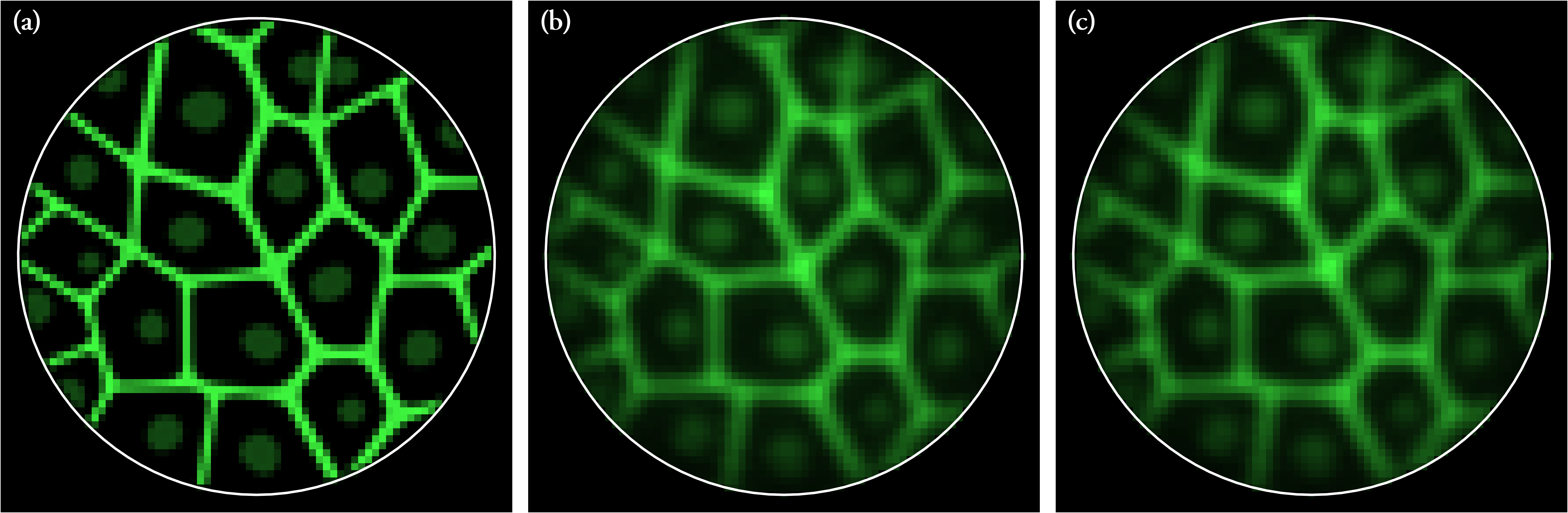}
\caption{{\bf Imaging with mode loss} {\bf (a)} An object at the distal end of the fibre, {\bf (b)} an image obtained with a lossless fibre, and {\bf (c)} an image obtained with a lossy fibre.}
\label{Fig:modeLoss}
\end{figure}

Under realistic experimental conditions fibre modes experience some power loss when propagating through a fibre, which could potentially reduce imaging quality. The energy carried by PIMs propagating with k-vectors closer to the angle of total internal reflection is more likely to escape through the cladding due to scattering from imperfections. We investigate this effect by imaging an object located at the distal end of the fibre, shown in Fig.\,\ref{Fig:modeLoss}(a). This object was chosen as it excites the majority of the 403 PIMs in the fibre. We simulate imaging this object through a lossy but otherwise ideal (no mode coupling) fibre. The result is shown in Fig.\,\ref{Fig:modeLoss}(c), and is virtually indistinguishable from the image obtained with an ideal lossless fibre, shown in Fig.\,\ref{Fig:modeLoss}(b). We note that the inverter itself also exhibits mode dependent loss, as higher order PIMs are less efficently sorted, the effect of which is present in all our simulated images. We expect that in practice, mode dependent loss inherent in the inverter will dominate over MMF mode dependent loss, as is the case here.

In our model, mode loss is a number in the range from zero to one, and it scales logarithmically with the phase velocity of each mode. This means that the higher order modes (those with the lowest phase velocity) experience more loss when propagating through the fibre, as is observed in experiments.

\newpage

\section{\S9: Combining images from different isoplanatic patches}
\begin{figure}[H]
\centering
\includegraphics[width=16cm]{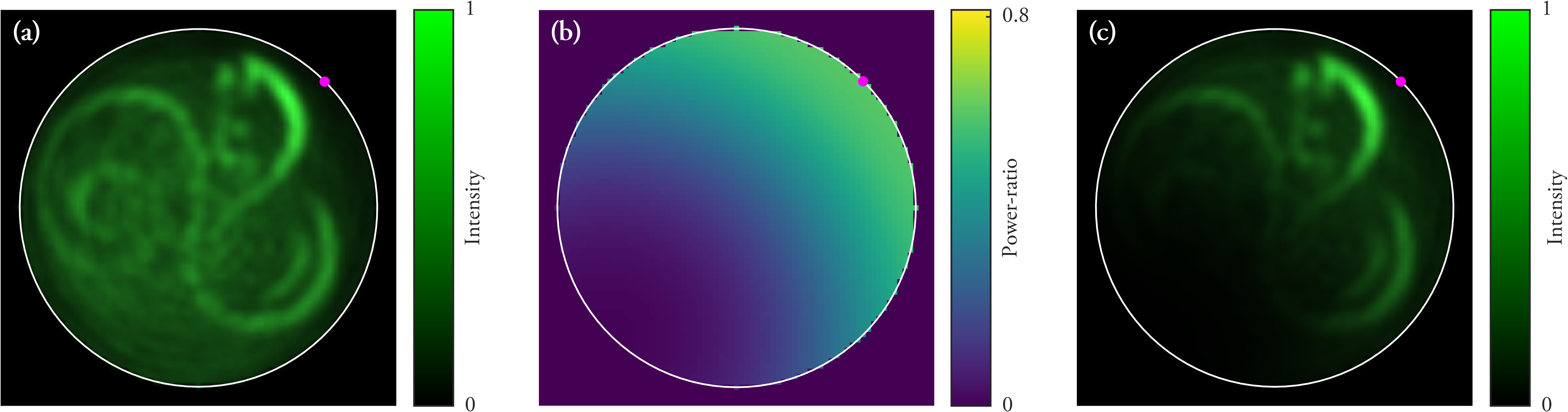}
\caption{{\bf Guide-star assisted imaging.} \textbf{(a)} An image obtained with the inverter using a guide-star (indicated with the pink dot), \textbf{(b)} an estimated power-ratio map, and \textbf{(c)} the element-wise product of the two (note that the intensity here has been rescaled to the maximum of 1).}
\label{fig:gs_rec}
\end{figure}

As discussed in the main paper, using a guide-star deployed at the distal end of the fibre can locally increase imaging contrast (when a non-ideal fibre is used). Several images obtained by scanning through different guide-star locations can then be averaged to recover the original scene. Taking an average directly, though, would result in a lot of noise, so we weight the images with estimated power-ratio maps that indicate the isoplanatic patch. This way low-contrast information is ignored and high contrast areas are revealed, as can be seen in Fig.~\ref{fig:gs_rec}.

If we let $A$ denote an image obtained with a guide-star, and $C$ denote the corresponding estimated power-ratio map, the reconstructed image is given by the weighted average:
\begin{equation}
    A_\text{rec} = \frac{1}{N}\sum_{i-1}^N A_iC_i,
\end{equation}
where $N$ is the total number of images, indexed by $i$.

Without the knowledge of the full TM of the fibre, the power-ratio map has to be approximated. Here we use a 2D Gaussian distribution, the minimum of which is centred on a point located at the edge of the fibre core directly opposite the guide star, and defined in cylindrical coordinates:
\begin{equation}
    C_j = \exp{\left(-\frac{(r_j-2r_0)^2}{2\sigma_r^2} - \frac{(\phi_j-\phi_0)^2}{2\sigma_\phi^2} \right)},
\end{equation}
where $j$ indexes the pixels, $r_0$ is the radial distance from the centre of the fibre core to the guide-star, $\phi_0$ is the azimuthal location of the guide-star, $r_j, \phi_j$ are the radial and azimuthal location of the $j$-th pixel respectively, and $\sigma_r, \sigma_\phi$ are the standard deviations, which can be varied to get the best estimate.

\newpage

\section{\S10: Description of Supplementary movies}

Movie 1: scanning a diffraction limited spot across the fibre facet. Here the first row illustrates incoherent imaging (the images show intensity), and the second row coherent imaging (the plots show complex amplitude of the fields).\\

Movie 2: super-resolution imaging. The leftmost panel shows an object at the distal end of the fibre (grey spiral), which is labelled with fluorophores. In every frame a different random subset of the fluorophores is excited to emit light (green dots). The signal is then imaged through our fibre-inverter system to form diffraction limited spots (middle panel). A region of interest can then be defined around each spot and its centre of mass tracked. The tracked centre-of-mass locations are than indicated in the third panel.\\

Movie 3: contracting isoplanatic patch. This video shows how modal cross-talk affects imaging capabilities through a fibre. As cross-talk increases ($p_\text{d}$ goes down) imaging contrast goes down but can be partially recovered with the use of guide star (magenta dot). The left panel shows a power-ratio map which indicates the imaging contrast, and the right panel shows an example object being imaged.

\end{document}